\documentclass[fleqn,usenatbib]{mnras}
\pdfoutput=1
\usepackage{newtxtext,newtxmath}

\usepackage{geometry}
\usepackage{float}
\usepackage{titlesec}
\usepackage{amsmath}
\usepackage[]{graphicx} \graphicspath{ {Fig/} }
\usepackage[]{xcolor}
\usepackage{booktabs}

\usepackage{epstopdf}
\usepackage{multicol}
\usepackage{graphicx}
\usepackage{subcaption}

\usepackage[T1]{fontenc}

\DeclareRobustCommand{\VAN}[3]{#2}
\let\VANthebibliography\thebibliography
\def\thebibliography{\DeclareRobustCommand{\VAN}[3]{##3}\VANthebibliography}

\usepackage{graphicx}	
\usepackage{amsmath}	

\title[Selection functions of lens finding NNs]{Selection functions of strong lens finding neural networks}
\author[A. Herle, C. M. O'Riordan and S. Vegetti]{
A. Herle,$^{1,2}$\thanks{E-mail: aherle@mpa-garching.mpg.de}
C. M. O'Riordan$^{1}$
and S. Vegetti$^{1}$
\\
$^{1}$Max Planck Institute for Astrophysics, Karl-Schwarzschild-Straße 1, 85748 Garching bei München, Germany\\
$^{2}$Universitäts-Sternwarte, Fakultät für Physik, Ludwig-Maximilians Universität München, Scheinerstr. 1, 81679 München, Germany\\}

\date{Accepted XXX. Received YYY; in original form ZZZ}

\pubyear{2023}

\begin{document}
\label{firstpage}
\pagerange{\pageref{firstpage}--\pageref{lastpage}}
\maketitle

\begin{abstract}
Convolution Neural Networks trained for the task of lens finding with similar architecture and training data as is commonly found in the literature are biased classifiers. An understanding of the selection function of lens finding neural networks will be key to fully realising the potential of the large samples of strong gravitational lens systems that will be found in upcoming wide-field surveys. We use three training datasets, representative of those used to train galaxy-galaxy and galaxy-quasar lens finding neural networks. The networks preferentially select systems with larger Einstein radii and larger sources with more concentrated source-light distributions. Increasing the detection significance threshold to 12$\sigma$ from 8$\sigma$  results in 50 per cent of the selected strong lens systems having Einstein radii $\theta_\mathrm{E}$ $\ge$ 1.04 arcsec from $\theta_\mathrm{E}$ $\ge$ 0.879 arcsec, source radii $R_S$ $\ge$ 0.194 arcsec from $R_S$ $\ge$ 0.178 arcsec and source S\'ersic indices $n_{\mathrm{Sc}}^{\mathrm{S}}$ $\ge$ 2.62 from $n_{\mathrm{Sc}}^{\mathrm{S}}$ $\ge$ 2.55. The model trained to find lensed quasars shows a stronger preference for higher lens ellipticities than those trained to find lensed galaxies. The selection function is independent of the slope of the power-law of the mass profiles, hence measurements of this quantity will be unaffected. The lens finder selection function reinforces that of the lensing cross-section, and thus we expect our findings to be a general result for all galaxy-galaxy and galaxy-quasar lens finding neural networks.
\end{abstract}




\section{INTRODUCTION}

Strong gravitational lensing allows one to address a broad range of cosmological and astrophysical questions, from the nature of dark matter \citep[e.g][]{Vegetti2018, Ritondale2019, Gilman2020, Hsueh2020} and galaxy evolution \citep[e.g.][]{Sonnenfeld2019, Mukherjee2021, Rizzo2021} to measuring the Hubble constant and other cosmological parameters \citep[e.g.][]{Birrer2021, Collett2014}. The field is set to profit enormously from ongoing and upcoming large-sky surveys (with e.g. Euclid, the Square Kilometre Array and the Vera Rubin Observatory), as the number of known strong gravitational lens systems is expected to increase by many orders of magnitude \citep{Collett2015, McKean2015}. An understanding of the selection function of these surveys is required for a correct interpretation of the scientific results from strong gravitational lensing analyses of the larger sample.

There are two sides to the problem of selection bias in the newly discovered strong lens systems: the effect of the lensing cross-section, and that of the lens finding method used to identify the lenses. \citet{Sonnenfeld2023} focused on the former and showed that strong lenses are a biased subset of the true population with respect to the lens and source parameters. The effect of the lens finding methodology, however, is still not well understood.

The task of finding strong gravitational lens systems amongst the large number of objects imaged in future surveys is a challenging one, especially as strong lenses are, by their nature, very rare. Hence, automation in lens finding is expected to play an important role in the identification of these systems.

Neural networks are a class of machine learning models that are often applied in astronomy to search for rare objects. Lens finder neural networks are those trained to sift through large volumes of image data specifically to find strong gravitational lens candidates. Over the years, Convolution Neural Networks (CNNs) have emerged as the state of the art for lens finding. These models are essentially classifiers, trained to distinguish between \emph{lens} and \emph{non-lens} gravitational systems. For example, \citet{Lanusse2018} developed CMU DeepLens, which is a CNN trained to find strong gravitational lens systems on LSST data. The ResNet architecture implemented in their work became the standard for this task, and was adapted for the DESI Legacy survey by \citet{Huang2020} and \citet{Huang2021}. CNNs have also been used to find strong lenses in the Kilo-Degree Survey \citep{Petrillo2017, Petrillo2019a, Petrillo2019b}, HST/ACS data of the COSMOS field \citep{Pourrahmani2018}, PANSTARS \citep{Canameras2020}, HSC \citep{Canameras2021, Canameras2023} and LOFAR \citep{Rezaei2022}.

In this paper, we present the first systematic study of the strong lens finder selection function. We develop lens finder neural networks in a similar fashion as commonly done in the lens finding literature and then characterise the selection effects that they introduce into the recovered sample of strong gravitational lenses. In particular, we are interested in identifying the characteristics of a strong gravitational lens system that drive the classification. Finally, we discuss the implications of these selection effects for different scientific applications of strong gravitational lensing.

The paper is organised as follows. In Section \ref{sect:formalism} we discuss the mathematical formalism used. In Section \ref{sect:data} we describe the image simulation process for the three datasets used in this work. In Section \ref{sect:methods} we detail the machine learning training and interpretability framework. In Section \ref{sect:results} we present our results on the selection function. In Section \ref{sect:disc} and \ref{sect:conc} we discuss our results and summarise our conclusions.
\section{SELECTION BIAS FORMALISM}
\label{sect:formalism}

From \citet{Sonnenfeld2023}, the probability $\mathrm{P}_{\mathrm{SL}}$ of selecting a sample of strong lens systems for a given selection criterion S can be expressed as:
\begin{equation}
\label{eqn:Psl}
    \mathrm{P}_{\mathrm{SL}}(\Psi_l, \Psi_s | S) \propto \mathrm{P}_l(\Psi_l) \mathrm{P}_s(\Psi_s)\mathrm{P}_{\mathrm{sel}}(\Psi_l, \Psi_s | S)\,.
\end{equation}
Here, $\Psi_l$ and $\Psi_s$ are the set of parameters describing the lens galaxy mass and the background source light distributions, respectively. $\mathrm{P}_l$ and $\mathrm{P}_s$ are the corresponding probability density distributions. 
$\mathrm{P}_{\mathrm{sel}}$ encapsulates the probability that a specific combination of lens and source produce a strong lens system and that this is found in the survey. It can be further separated into two components:
\begin{equation}
\mathrm{P}_{\mathrm{sel}}(\Psi_l, \Psi_s | S) = \mathrm{P}_{\mathrm{det}}(\Psi_l, \Psi_s)\mathrm{P}_{\mathrm{find}}(\Psi_l, \Psi_s | S).
\end{equation}
where $\mathrm{P}_{\mathrm{det}}$ is the probability that multiple images of the source form as clearly distinct features in the survey image, and $\mathrm{P}_{\mathrm{find}}$ is the probability that this image is correctly classified as a strong lens system. The latter depends on the identification procedure adopted. By focusing on the first term, \citet{Sonnenfeld2022} and \citet{Sonnenfeld2023} have quantified how strongly the properties ($\Psi_l$ and $\Psi_s$) of the lenses and background sources are biased with respect to the general population of galaxies. The goal of this paper is to quantify $\mathrm{P}_{\mathrm{find}}$ for the case in which a CNN is used to identify lens systems in a given survey.
\begin{figure*}
    \centering
    \includegraphics[scale=0.5]{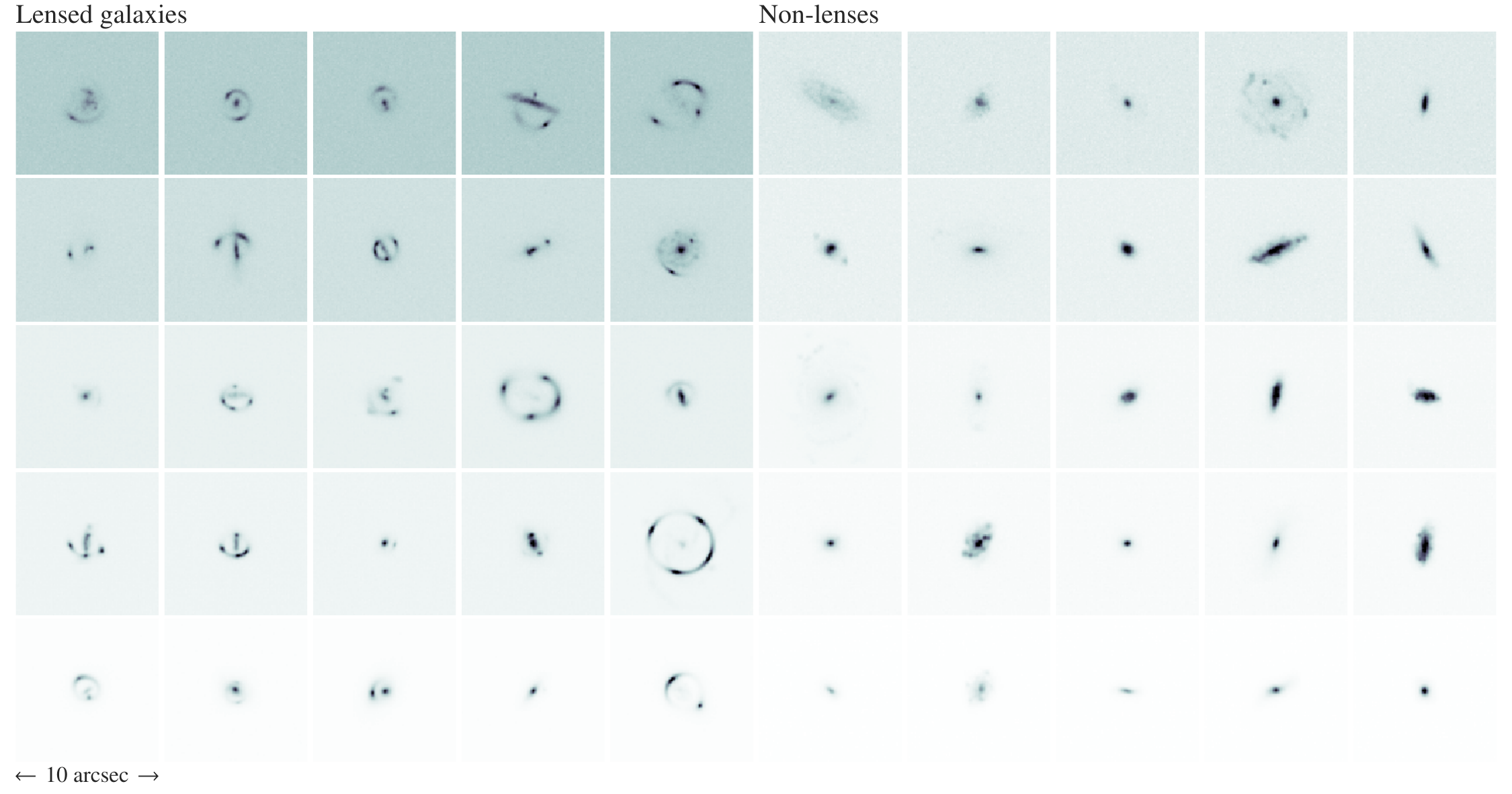}
    \caption{A representative sample from $\mathcal{D_B}$, ranked in increasing order of SNR from the top left to the bottom right in each panel. Note the range of complexities of the lens light models in the sample. In some cases, distinguishing between lens light and lensed source emission is trivial, but in cases where the lens light model contains complex features like spiral arms, this is not the case. Examples of lenses are shown on the left, and examples of non-lenses are shown on the right. The simulation parameters are sampled as shown in Table \ref{table:dataset_params}.}
    \label{fig:datasetB}
\end{figure*}
\begin{figure*}
    \centering
    \includegraphics[scale=0.5]{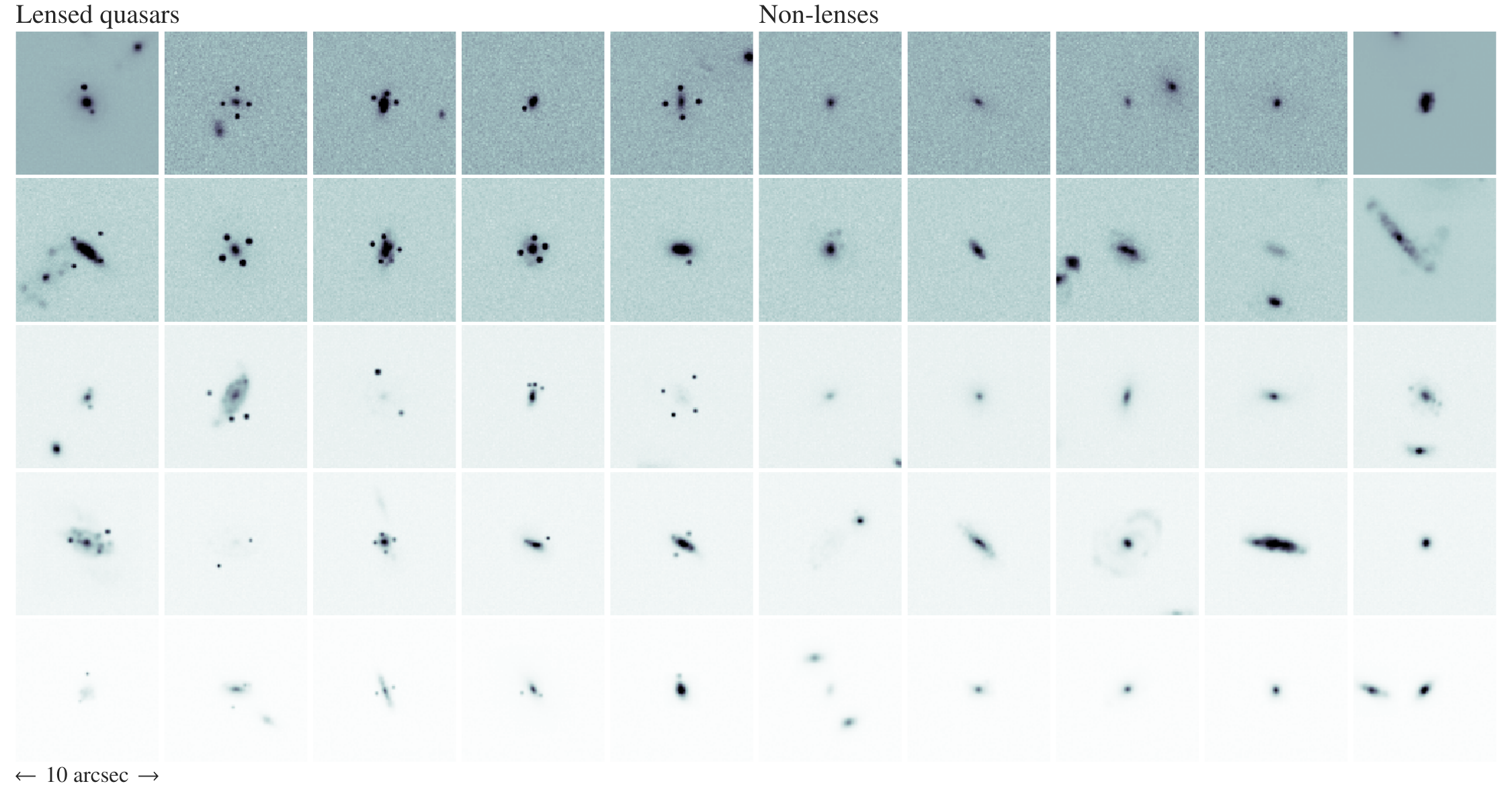}
    \caption{A representative sample from $\mathcal{D_C}$, ranked in increasing order of SNR from the top left to the bottom right in each panel. The images are simulated as described in Section \ref{sect:methods}, with the inclusion of contaminant galaxies and a variable PSF FWHM. Examples of lenses are shown on the left, and examples of non-lenses are shown on the right. The simulation parameters are sampled as shown in Table \ref{table:dataset_params}.}
    \label{fig:datasetC}
\end{figure*}
\section{DATA}
\label{sect:data}

\begin{table*}
\caption{Parameter distributions used to create all datasets. From top to bottom: the parameters describing the source light distribution, the mass and light properties of the lens, distribution for the light contaminants and the observational set up. The superscripts S and L denote source and lens properties respectively.} 
\centering
\begin{tabular}{l |c|c|c|c|} 
\hline
  Parameter & $\mathcal{D_A}$ & $\mathcal{D_B}$ & $\mathcal{D_C}$\\
  \hline
  Source Light & S\'ersic & S\'ersic & Point-like\\
  Source Radius, $R_{\mathrm{S}}$ (arcsec) & $\mathcal{U}(0.05, 0.3)$ & $\mathcal{U}(0.05, 0.3)$ & -\\
  Source S\'ersic Index, $n_{\mathrm{Sc}}^{\mathrm{S}}$ & $\mathcal{U}(1, 4)$ & $\mathcal{U}(1, 4)$ & -\\
  Source Axis-Ratio, $q_{\mathrm{S}}$ & $\mathcal{U}(0.5, 1.0)$ & $\mathcal{U}(0.5, 1.0)$ & -\\
  Source Redshift & 2.0 & 2.0 & 2.0 \\
  Source Apparent Magnitude, $M_{\mathrm{VIS}}^{\mathrm{S}}$ & $\mathcal{U}(M_{\mathrm{VIS}}^{\mathrm{L}}, 24.0)$ & $\mathcal{U}(M_{\mathrm{VIS}}^{\mathrm{L}}, 24.0)$ & - \\
  Source Flux & - & - & 1.0 \\
  \\
  Lens light & S\'ersic & GAN & GAN\\
  Lens mass axis-ratio, $q$ & $\mathcal{U}(0.5, 1.0)$ & - & -\\ 
  Lens power-law slope, $\gamma$ & $\mathcal{U}(1.8, 2.2)$ & $\mathcal{U}(1.8, 2.2)$ & $\mathcal{U}(1.8, 2.2)$ \\ 
  Einstein Radius, $\theta_\mathrm{E}$ (arcsec)& $\mathcal{U}(0.5, 2.0)$ & - & - \\
  Lens redshift & 0.8 & 0.8 & 0.8\\
  Lens S\'ersic Index, $n_{\mathrm{Sc}}^{\mathrm{L}}$ & $\mathcal{U}(1, 4)$ & - & -\\
  Lens Apparent Magnitude, $M_{\mathrm{VIS}}^{\mathrm{L}}$ & $\mathcal{U}(18.0, 22.0)$ & $\mathcal{U}(18.0, 22.0)$ & -\\
  Lens Flux, $F_\mathrm{L}$  & - & - & $\mathcal{U}(5.0, 100.0)$ \\
  Shear Strength & $\mathcal{U}(0.0, 0.1)$  & $\mathcal{U}(0.0, 0.1)$ & $\mathcal{U}(0.0, 0.1)$\\
  \\
  Contaminants & No & No & Yes\\
  Contaminant Flux, $F_\mathrm{C}$ & - & - & $\mathcal{U}(1.0, 100.0)$  \\
  \\
  Pixel Size (arcsec) & 0.1 & 0.1 & 0.1 \\
  Field-of-View (arcsec) & 10 & 10 & 10 \\
  $\mathrm{PSF}_{\mathrm{fwhm}}$ (arcsec) & 0.16 & 0.16 & $\mathcal{U}(0.16, 0.3)$ \\
  \hline
\end{tabular}
\label{table:dataset_params}
\end{table*}

The Euclid mission is expected to find about $10^5$ strong gravitational lens systems \citep{Collett2015}. In this work, we focus on neural networks trained only with single-band data because most of the gravitational lens systems that will be found in the Euclid Wide Survey will be observed using the Visual imager (VIS) instrument. 

VIS is a broadband optical instrument with a resolution of 0.16 arcsec, which is about three times better than that of the three infrared bands of the Near Infrared Spectrometer and Photometer (NISP) instrument \citep{O'Riordan2023, Euclid2022}. Since the Einstein radius of the strong gravitational lenses expected to be found by Euclid peaks around 0.5 arcsec \citep{Collett2015}, the arcs and lens light are expected to be blended together within the NISP instrument. Hence, these features will be much better resolved by VIS, at the cost of losing colour information. The absence of multi-band data can be crucial in informing the decision rationale that CNNs arrive at during training, and may have a significant effect on the type of strong gravitational lens systems that will be identified. Moreover, \citet{Petrillo2019a} found that the best performing network in terms of purity and completeness was the one trained on single-band data. Similarly, \citet{Lanusse2018} showed that CMU DeepLens is a successful tool for finding strong gravitational lens systems, and that even without colour information, it is able to learn enough about the lensed-arc morphology to solve the classification problem.

We make three datasets, $\mathcal{D_A}$, $\mathcal{D_B}$ and $\mathcal{D_C}$, with $10^6$ images for training and $2 \times 10^5$ for testing each. The samples are split evenly between two distinct classes, \emph{lens} and \emph{non-lens}, which contain all and none of the lens systems, respectively.

Fig. \ref{fig:datasetB} and \ref{fig:datasetC} show examples of the two classes from $\mathcal{D_B}$ and $\mathcal{D_C}$, and Table \ref{table:dataset_params} summarises the simulation parameter distributions for all datasets.

We refer the reader to the following sections for more details on the properties of each dataset. Briefly, the datasets $\mathcal{D_A}$ and $\mathcal{D_B}$ have extended sources, while the sources in $\mathcal{D_C}$ are unresolved. This is done to understand how the selection function, $\mathrm{P}_{\mathrm{find}}$, differs between galaxy-galaxy and galaxy-quasar lens systems. The lens galaxies in $\mathcal{D_A}$ have a simple analytical model for the lens-light distribution, while those in $\mathcal{D_B}$ and $\mathcal{D_C}$ are characterised by a higher level of complexity. This is done in order to quantify how the lens light model complexity affects the neural network selection function.

We add complexity to the lens rather than the source light because distinguishing between these components becomes non-trivial when the lens contains features resembling lensed source emission (e.g. arcs). The lensed  emission is typically sufficiently distinct from analytic S\'ersic components because of its distorted morphology, such that introducing structure in the source will only make the task of lens finding easier. Moreover, using analytical source models allows us to more easily quantify how their parameters affect the network selection function for different choices of the lens light distribution.

\subsection{Source light}

We use S\'ersic profiles \citep{Sersic1963, Ciotti1999} for the source-light model in the case of $\mathcal{D_A}$ and $\mathcal{D_B}$. A point-like source model is used for $\mathcal{D_C}$. We refer the reader to Table \ref{table:dataset_params} for more detailed information.

\subsection{Lens light}

The lens light distribution in $\mathcal{D_A}$ is created using a S\'ersic profile  with an index and effective radius sampled uniformly from between 1.0 and 4.0 and between 0.05 and 0.5 arcsec, respectively. $\mathcal{D_B}$ and $\mathcal{D_C}$ use images generated with a Generative Adverserial Network \citep[GAN, see][for more details]{Holzschuh2022}. The GAN was trained to generate images that imitate those created from the SKIRT code on the IllustrisTNG simulation \citep{Springel2018, Rodriguez-Gomez2019}. The resulting datasets consist, therefore, of realistic early- (i.e. S\'ersic-like) and late-type galaxy images, several of which contain complex features like star-forming clumps, spiral arms and satellite galaxies (see Fig. \ref{fig:datasetB} for examples of lens systems taken from $\mathcal{D_B}$). Structures of this kind can easily be confused for lensed-source emission when dealing with single-band data, making the problem of lens-finding more challenging. In this respect, $\mathcal{D_A}$ represents the simplest formulation of the problem for the network to solve. 

The lens magnitudes in $\mathcal{D_A}$ and $\mathcal{D_B}$ are sampled from a uniform distribution, and the source magnitudes are chosen such that they are dimmer than the lens. In $\mathcal{D_C}$, we instead use fluxes defined relative to the source. In practice, the lens flux is set such that it is 5 to 100 times brighter (sampled uniformly in this range) than the source light. We chose these values based on the distribution of magnifications of the lens mass models (see Section \ref{sec:lens_mass} for more details) in this dataset, which peaks at $\mu \approx 3$, thus ensuring that the number of systems where the lens light is completely absent in the image are negligible. These values are also chosen such as to keep the flux per pixel area the same, as the source light is concentrated into a point.

\subsection{Light contaminants}
\label{sec:contaminants}

We do not include any light contaminants in $\mathcal{D_A}$ and $\mathcal{D_B}$. On the other hand, $\mathcal{D_C}$ does contain light contribution from nearby field galaxies. These are sampled from the GAN dataset and are placed randomly on the image, while ensuring that they lie outside the Einstein radius of the system and at least a set minimum distance apart from each other. These contaminants are 1 to 100 times brighter than the source light and their number in each image is sampled uniformly between 0 and 4. We sample their redshift between 0.2 and 5.0 from a probability density distribution based on the comoving volume at these redshifts. Examples of gravitational lens systems from $\mathcal{D_C}$ are shown in Fig. \ref{fig:datasetC}.

\subsection{Lens mass}
\label{sec:lens_mass}
For all three datasets, an elliptical power law is assumed for the lens mass model with the slope, $\gamma$, ranging uniformly between 1.8 and 2.2 \citep{Koopmans2006}. In the case where the lens light is a GAN image (i.e. $\mathcal{D_B}$ and $\mathcal{D_C}$), we use the image moments of the latter to align the position angle and axis-ratio of the mass profile with those of the lens light, thus ensuring that the light and mass distributions are consistent with each other. Moreover, the Einstein radius is scaled proportionally to the total flux of the selected GAN image. Thus, $\mathcal{D_B}$ and $\mathcal{D_C}$ have the same simulation parameters as $\mathcal{D_A}$, except for the Einstein radius and lens axis-ratio, for which the distributions are shown in the left and centre panels of Fig. \ref{fig:datasetB_parameters}, respectively.

External shear with no preferred direction and with strength sampled from a uniform distribution is applied to all datasets. Additionally, for $\mathcal{D_C}$, we further ensure that the numbers of doubly- and quadruply-imaged systems are equal within the lens class. This is done by first calculating the area inside the inner caustic for the specific axis-ratio chosen. We then find the radius of a circle that would be twice this area and sample the source position uniformly from within this circle.
\subsection{Noise}

In order to add noise to the data, we follow the definition of signal-to-noise ratio (SNR) by \citet{O'Riordan2019}: 

\begin{equation}
\label{eqn:snr}
S_{\mathrm{T}} = \frac{\Sigma_i ^N m_i \mathrm{d}_i}{\sigma_d \sqrt{\Sigma_i ^N m_i}}\,,
\end{equation}

where $\sigma_d$ is the standard deviation of the sky noise, $\mathrm{d}_i$ is the pixel value at index $i$, and $m_i$ is the masking variable defined such that $m_i = 1$ for pixels corresponding to lensed source-light and 0 otherwise. Pixel values that lie within twice the Einstein radius of the lens subtracted image of the system are considered to be source-light for $\mathcal{D_A}$ and $\mathcal{D_B}$, and in the case of $\mathcal{D_C}$, this is considered to be all pixels with a value greater than 1.5 times the standard deviation of the lens subtracted image.

\begin{figure*}
    \centering
    \includegraphics[scale=0.7]{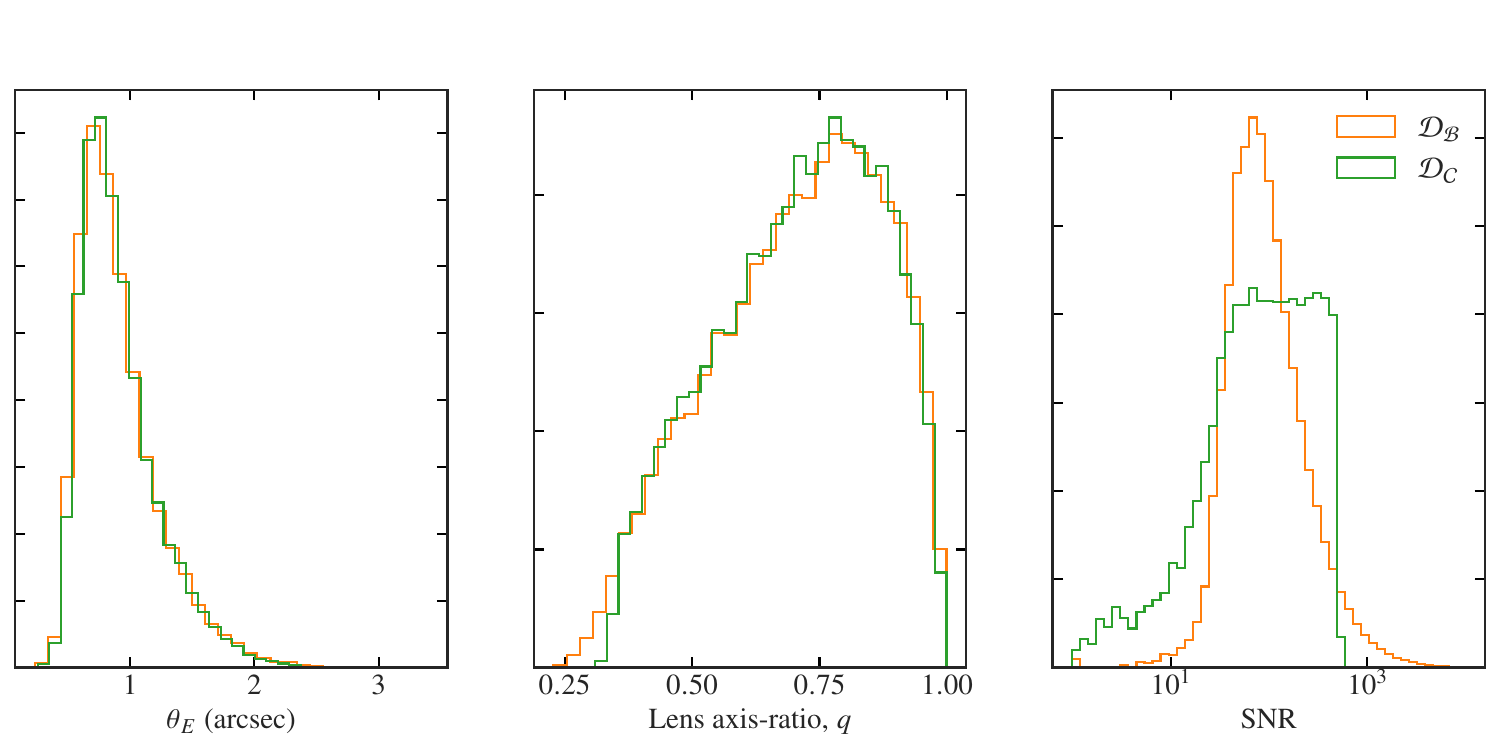}
    \caption{From left to right: distribution of lens Einstein radius, axis-ratio and SNR of the images for the datasets $\mathcal{D_B}$ (orange histograms) and $\mathcal{D_C}$ (green histograms). The bins are in log-space for the panel on the right.}
    \label{fig:datasetB_parameters}
\end{figure*}

The SNR of each image in $\mathcal{D_A}$ and $\mathcal{D_B}$ is set by the specific combination of source and lens magnitudes, and the resulting distribution is shown in the right panel of Fig. \ref{fig:datasetB_parameters}. In the case of $\mathcal{D_C}$, we first sample a value for the SNR from a uniform distribution (see Table \ref{table:dataset_params}), then add uncorrelated Gaussian noise to the images with a standard deviation calculated from Eqn. (\ref{eqn:snr}).

\subsection{Point-Spread Function}

We use a circular Gaussian function for the  Point-Spread Function (PSF). The Full Width at Half Maximum (FWHM) has a value of 0.16 arcsec for $\mathcal{D_A}$ and $\mathcal{D_B}$. For $\mathcal{D_C}$, we employ a variable FWHM, which is sampled uniformly between 0.16 and 0.3 arcsec. A FWHM of 0.16 arcsec corresponds to that of the VIS instrument aboard the Euclid telescope, and 0.3 arcsec represents one that is about 2 times worse than this. 

\section{METHODS}
\label{sect:methods}

\subsection{Training phase}

During the training phase, the datasets were split into batches of $10^3$ images each. ResNet18 CNNs were trained for 800 epochs each on $\mathcal{D_A}$ and $\mathcal{D_B}$. For $\mathcal{D_C}$, we also included dynamic learning rate scheduling, wherein the learning rate is decreased by a factor $10^{-0.5}$ if the test loss remains stagnant for 20 epochs. These values were chosen after extensive hyper-parameter tuning. The convergence criteria is set to be three consecutive drops in learning rate without a change in the test loss. With this setup, the network is trained for 260 epochs.

For all datasets, several data augmentation techniques were used during training, namely: random cropping to 80 $\times$ 80 pixels, random vertical and horizontal flips, rotation and erasing. The learning rate was set to 0.001 and the Adam optimizer was used to minimize the binary cross entropy loss. The images were normalised such that each pixel value is between 0 and 1 before being passed to the neural network.

Table \ref{table:datasetA_B-training} shows the differences in the final testing accuracy and loss for each of the three networks, as well as the True Positive Rate (TPR), False Positive Rate (FPR) and Area Under the Receiver Operator Characteristic (AUROC) curve for each network with corresponding dataset version. All three networks achieve very high classification accuracy, and have converged. This is also reflected in the AUROC values: the closer the AUROC is to 1, the closer the network is to a perfect classifier; thus the AUROC curve is a measure of the quality of the classifier for the particular dataset it is trained on.

The small decrease in network performance between $\mathcal{D_A}$ and $\mathcal{D_B}$ is a consequence of the fact that the network is exposed to a larger fraction of lens-light models and negative examples (galaxies in the \emph{non-lens} class) that have complex structure which can be confused for lensed arcs. We note that the higher test accuracy of the network trained on $\mathcal{D_A}$ does not imply a superior performance, rather, it indicates the relative simplicity of the lens light model.
\subsection{Detection significance of a \emph{lens} image}
We use the inverse error function to convert from probabilities to detection significance:
\begin{equation}
    C = \sqrt{2} \ \mathrm{erf}^{-1}(p)\,,
\end{equation}
where $C$ is the network lens detection significance and $p$ is the probability that an image belongs to the \emph{lens} class. More details can be found in the Appendix \ref{sect:logits}. In a lens finding campaign, a detection significance threshold is chosen such that all classifications made by the network above this $\sigma$-cut constitute a strong lens system. The distributions of the outputs for the three different networks on their respective datasets, after converting to significance, are shown in the right panel of Fig. \ref{fig:det-sig-distr}. Note that the distributions are roughly Gaussian and peak at different values for each dataset.

\subsection{Kullback-Leibler divergence}
\label{sect:KLD}
We would like to quantify the difference between the ground truth population of lenses and sources ($\Psi_l$, $\Psi_s$) and the population recovered by the lens finder. To do this, we employ the Kullback-Leibler (KL) divergence, which is defined as:
\begin{equation}
    D_{KL} (p||q) = - \int p(x) \ln \left [\frac{q(x)}{p(x)} \right] dx\,.
\end{equation}
Here, $p(x)$ is the probability distribution function of the true parent distribution (all lenses in the testing dataset) for the parameter $x$, and $q(x)$ is that of the selected sample at a particular detection significance threshold. We estimate these distribution functions at a specific $\sigma$-cut using Kernel Density Estimation (KDE) with a Gaussian kernel. We use Scott's rule \citep{Scott2015} to estimate a reasonable bandwidth for the KDE.
\begin{figure*}
    \centering
    \includegraphics[scale=0.7]{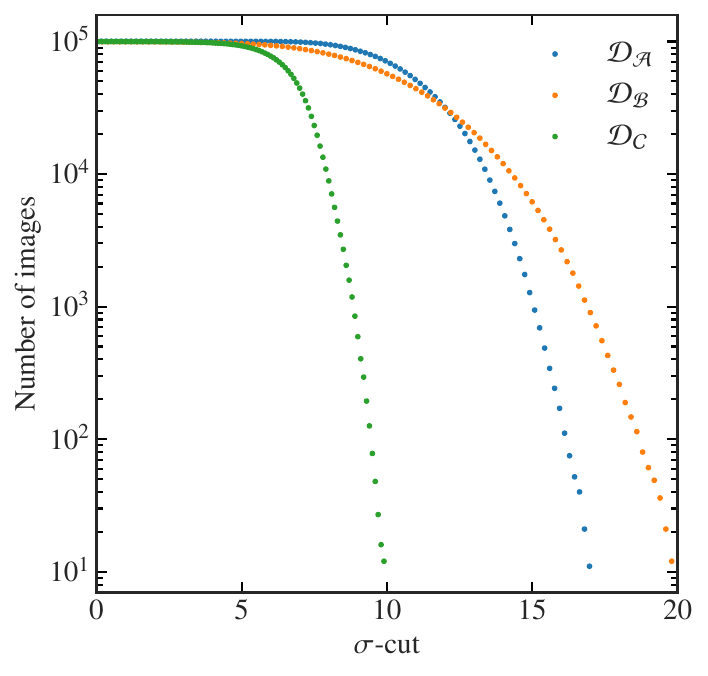}
    \hspace{-0.3cm}
    \includegraphics[scale=0.7]{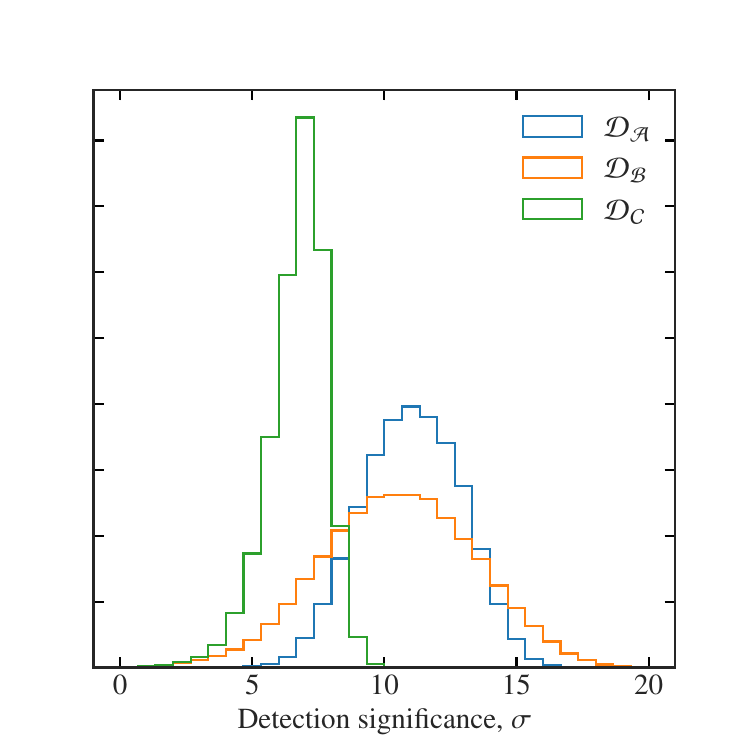}
    \caption{Left: the number of \emph{lens} images at each detection significance threshold for $\mathcal{D_A}$, $\mathcal{D_B}$ and $\mathcal{D_C}$. Right: distribution of detection significance values for the three networks trained on $\mathcal{D_A}$, $\mathcal{D_B}$ and $\mathcal{D_C}$.}
    \label{fig:det-sig-distr}
\end{figure*}

The purpose of using the KL-divergence as a metric is two-fold: (i) from an interpretability point of view, the parameters that show an increase in the KL-divergence at high detection significance thresholds are clearly important for the networks to classify a system as a \emph{lens} or a \emph{non-lens}, lending insight into what the networks have learnt in order to solve the lens finding problem. (ii) For physical parameters that are measured in a strong lensing survey, the KL-divergence indicates by how much the parameters inferred will deviate from the parent distribution for a given detection threshold. 

It must be noted here that we have neglected the correlations between the different parameters when calculating the KL-divergence, which is akin to marginalising over all parameters except the one being considered in the calculation. This means that the variation of the KL-divergence for a specific parameter as the detection significance threshold becomes stricter indicates by how much this parameter will differ from the true sample, if we were only interested in measuring this specific parameter in a given survey. A more thorough analysis, which is beyond the scope of this work, would require the covariance of the parameters to be taken into account.

We also note that as the detection significance threshold increases, the number of lens images drops exponentially, as shown in the left panel of Fig. \ref{fig:det-sig-distr}. Thus, there is an increase in the KL-divergence due to sampling noise, which needs to be accounted for in order to disentangle it from the increase in the KL-divergence that is a result of the selected sample differing from the truth. To this end, we sample the true distribution with different numbers of total samples (N) corresponding to each $\sigma$-cut, and then calculate the KL-divergence of this sample relative to the the entire true distribution of $10^5$ systems. This is repeated several thousands of times to get enough realisations to account for sampling variance. 

\begin{table}
\caption{Network performance in terms of the final test accuracy, test dataset loss, the True Positive and False Negative Rate, and the Area Under ROC curve, for each dataset.}
\label{table:datasetA_B-training}
\centering
\begin{tabular}{|c|c|c|c|c|c|} 
\hline
 Dataset & Test Accuracy & Loss & TPR & FPR & AUROC\\
 & [per cent] & & [per cent] & [per cent]\\
  \hline
  $\mathcal{D_A}$ & 99.99 & 0.035 & 99.99 & 0.002 & 0.9999998\\
  $\mathcal{D_B}$ & 99.64 & 0.064 & 99.61 & 0.202 & 0.9998687\\
  $\mathcal{D_C}$ & 99.88 & 0.006 & 99.89 & 0.081 & 0.9999866\\
  \hline
\end{tabular}
\end{table}
\section{RESULTS}
\label{sect:results}

\begin{figure*}
    \centering
    \includegraphics[scale=0.3]{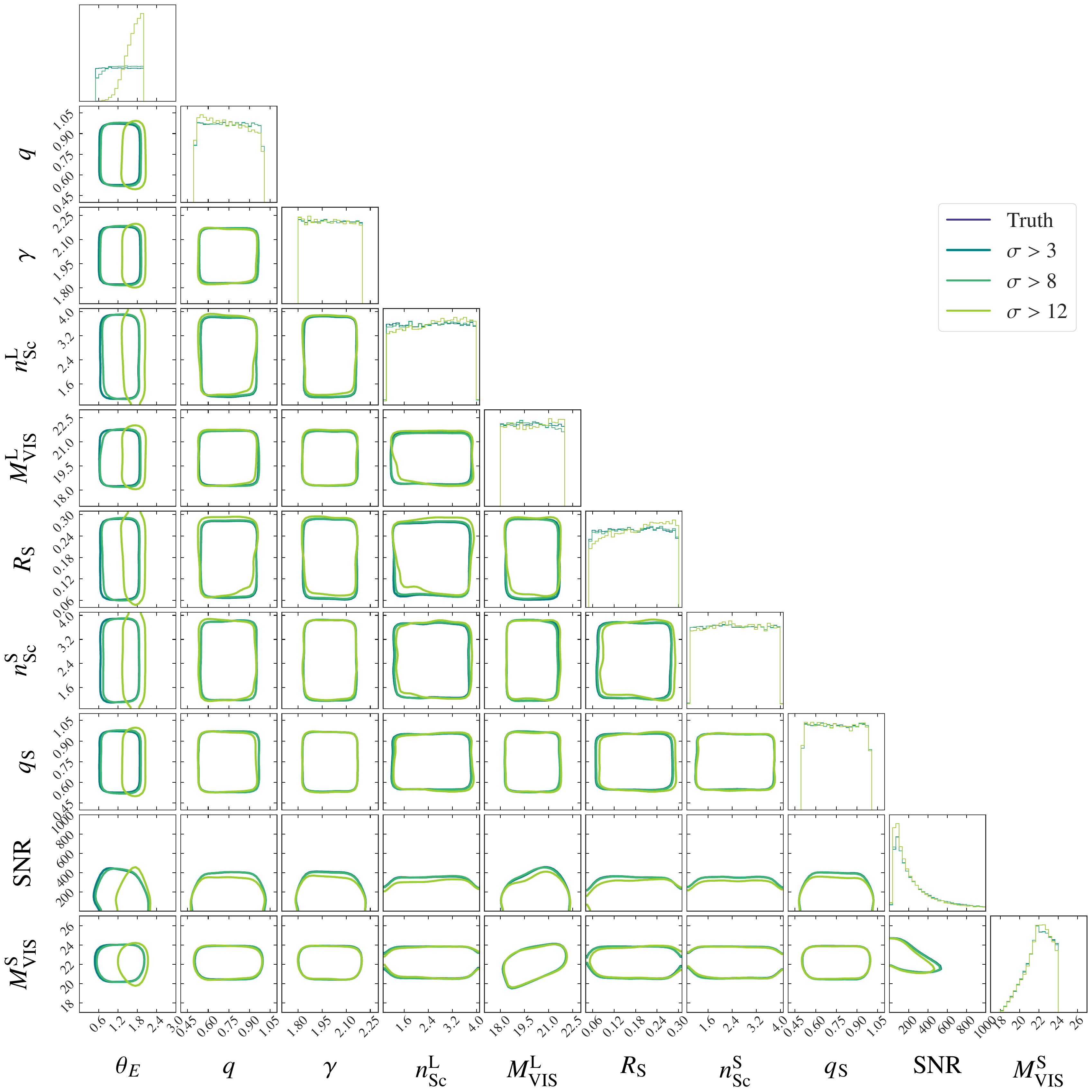}
    \caption{Distributions of the parent and selected sample for detection significance thresholds of 3, 7 and 10$\sigma$ for the network trained on $\mathcal{D_A}$. $\theta_\mathrm{E}$ is the Einstein radius, $q$ is the axis-ratio of the lens mass, $\gamma$ is the power-law slope, $M_{\mathrm{VIS}}^{\mathrm{L}}$ and $M_{\mathrm{VIS}}^{\mathrm{S}}$ are the apparent magnitude of the lens and source respectively, $R_{\mathrm{S}}$ is the source radius, $n_{\mathrm{Sc}}^{\mathrm{S}}$ is the S\'ersic index of the source, $q_{\mathrm{S}}$ is the axis-ratio of the source and SNR is the signal-to-noise ratio.}
    \label{fig:corner-sig-A}
\end{figure*}

\begin{figure*}
    \centering
    \includegraphics[scale=0.3]{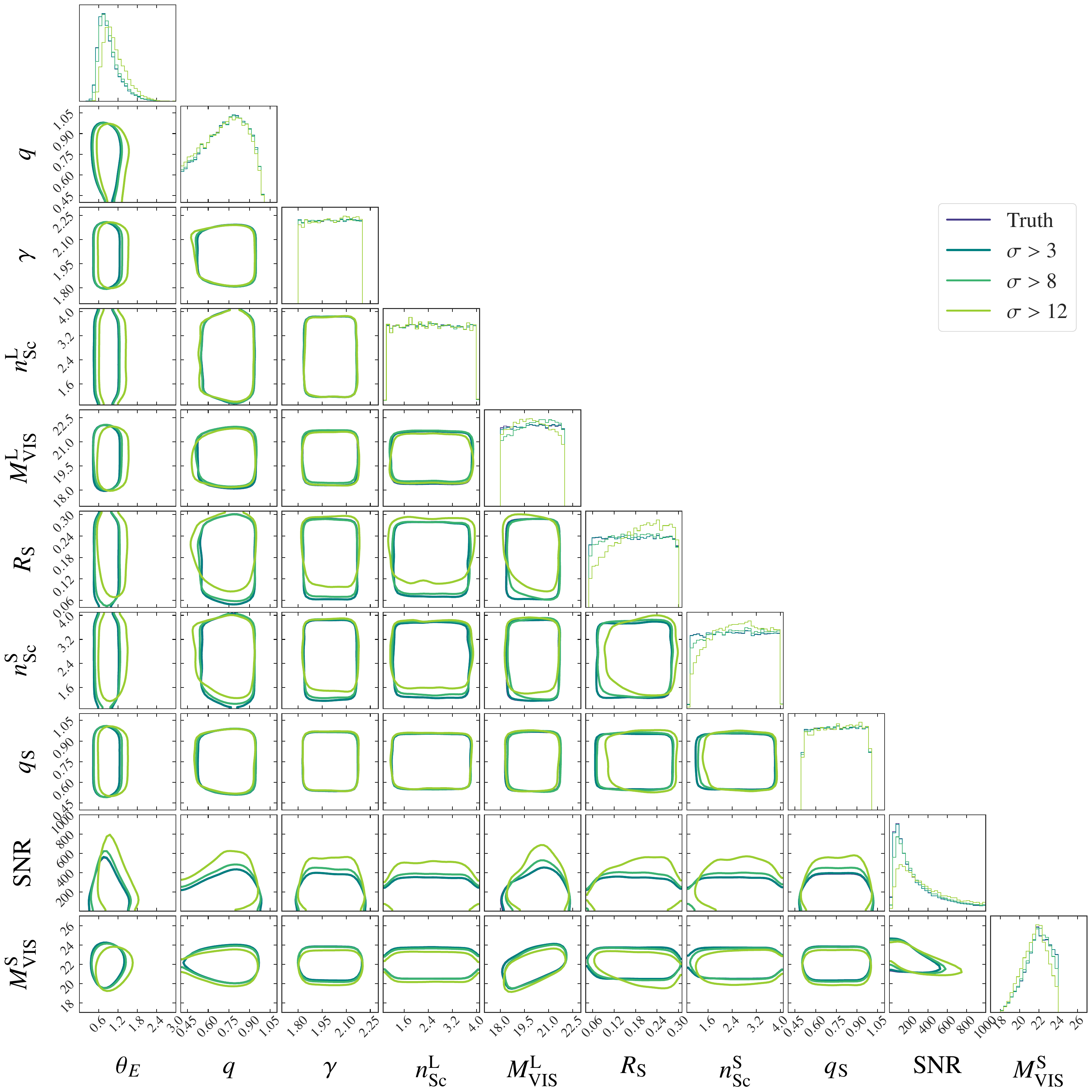}
    \caption{Distributions of the parent and selected sample for detection significance thresholds of 3, 7 and 10$\sigma$ for the network trained on $\mathcal{D_B}$.  $\theta_\mathrm{E}$ is the Einstein radius, $q$ is the axis-ratio of the lens mass, $\gamma$ is the power-law slope, $M_{\mathrm{VIS}}^{\mathrm{L}}$ and $M_{\mathrm{VIS}}^{\mathrm{S}}$ are the apparent magnitude of the lens and source respectively, $R_{\mathrm{S}}$ is the source radius, $n_{\mathrm{Sc}}^{\mathrm{S}}$ is the S\'ersic index of the source, $q_{\mathrm{S}}$ is the axis-ratio of the source and SNR is the signal-to-noise ratio.}
     \label{fig:corner-sig-B}
\end{figure*}

\begin{figure*}
    \centering
    \includegraphics[scale=0.35]{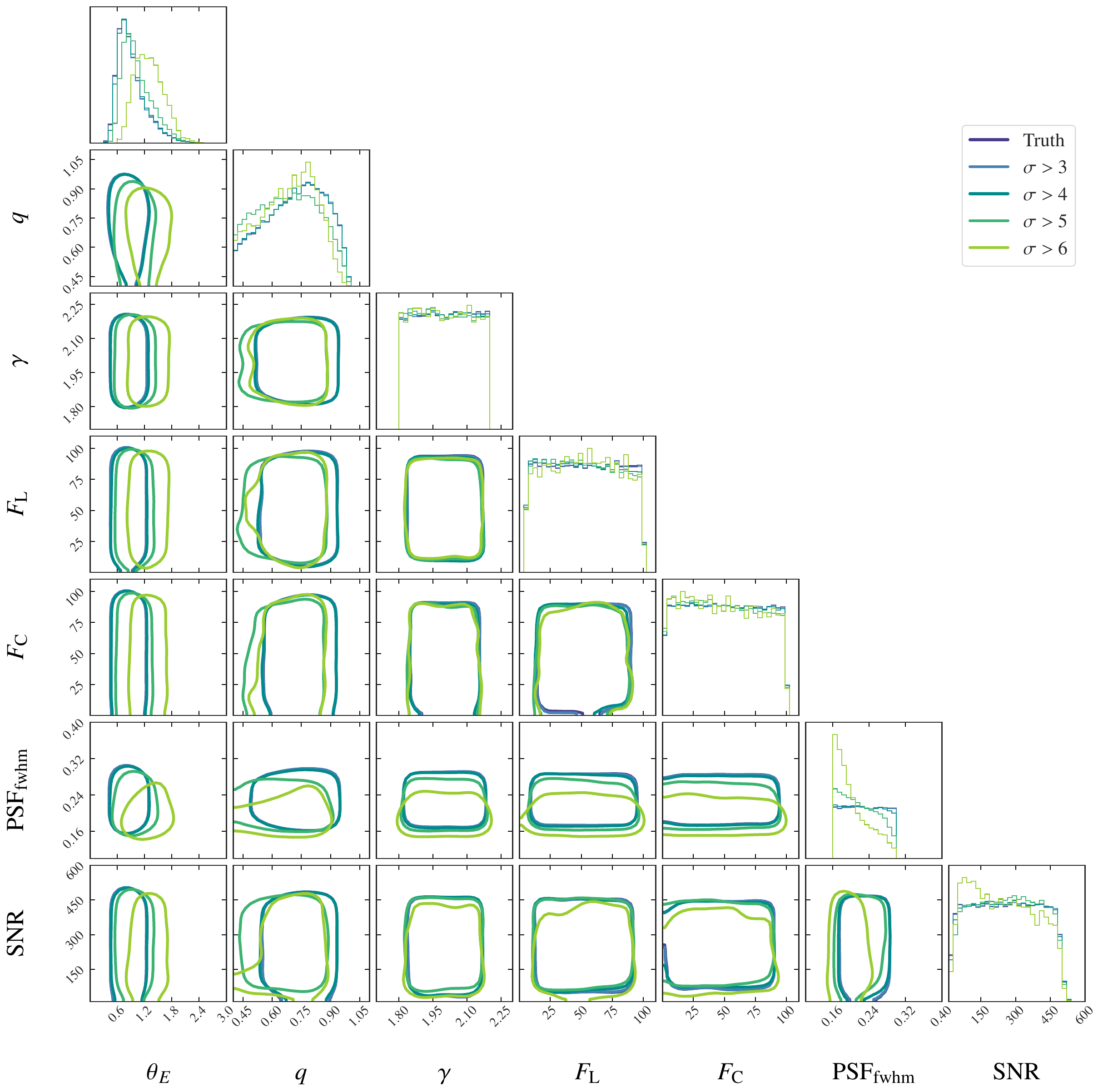}
    \caption{Distributions of the parent and selected sample for detection significance thresholds of 3, 4, 5 and 6$\sigma$ for the network trained on $\mathcal{D_C}$.  $\theta_\mathrm{E}$ is the Einstein radius, $q$ is the axis-ratio of the lens mass, $\gamma$ is the power-law slope, $F_{\mathrm{L}}$ and $F_{\mathrm{C}}$ are the lens and contaminant flux respectively and SNR is the signal-to-noise ratio.}
    \label{fig:corner-sig-C}
\end{figure*}

\begin{figure*}
    \centering
    \includegraphics[scale=0.38]{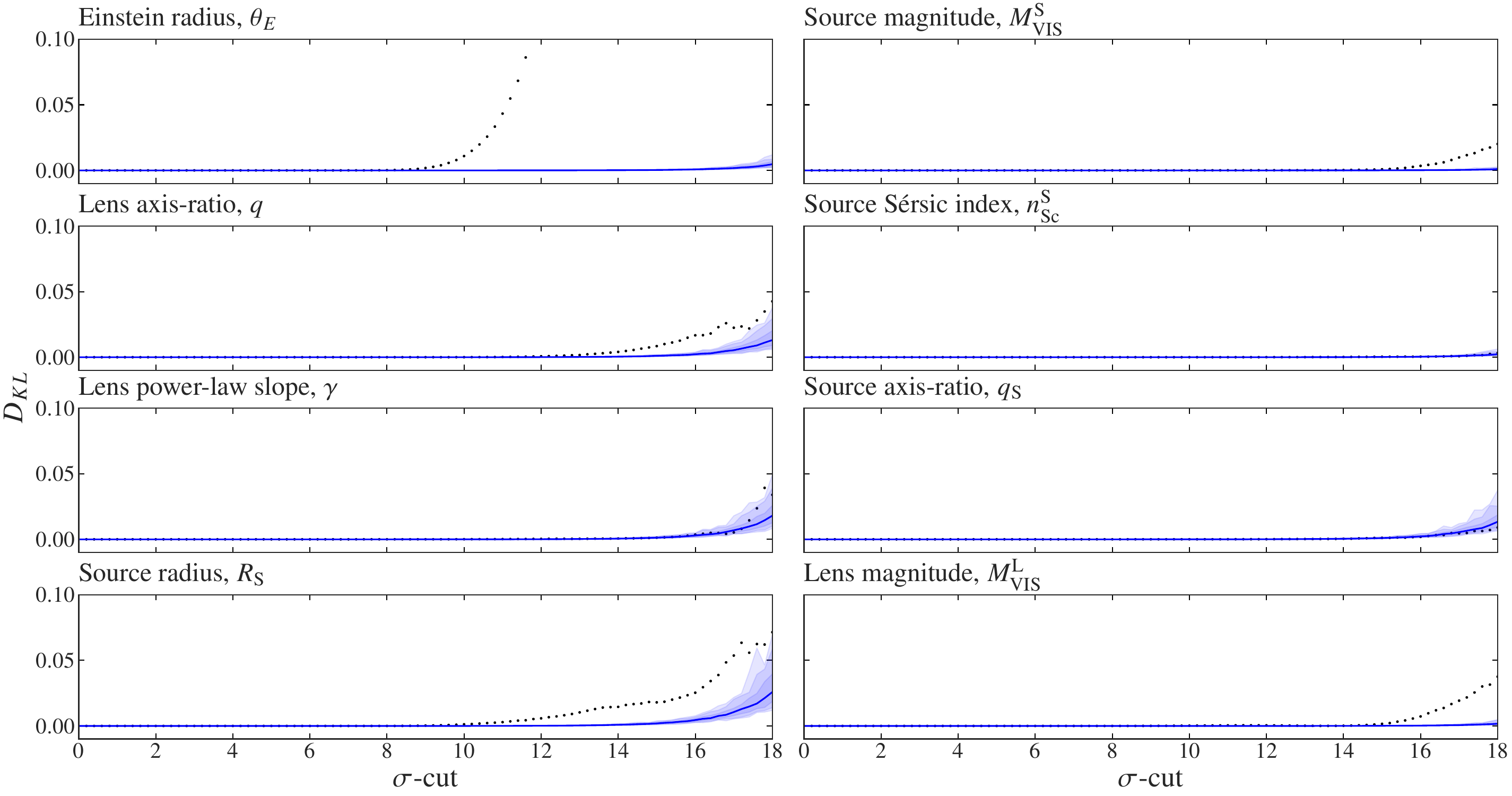}
    \caption{The increase in the Kullback-Leibler Divergence ($D_{KL}$) as the detection significance threshold is increased for the network trained on $\mathcal{D_A}$. The blue lines show the $D_{KL}$ that is due to sampling noise, with the increasingly shaded regions depicting 1, 2 and 3$\sigma$ contours.}
    \label{fig:dkl_A}
\end{figure*}

\begin{figure*}
    \centering
    \includegraphics[scale=0.38]{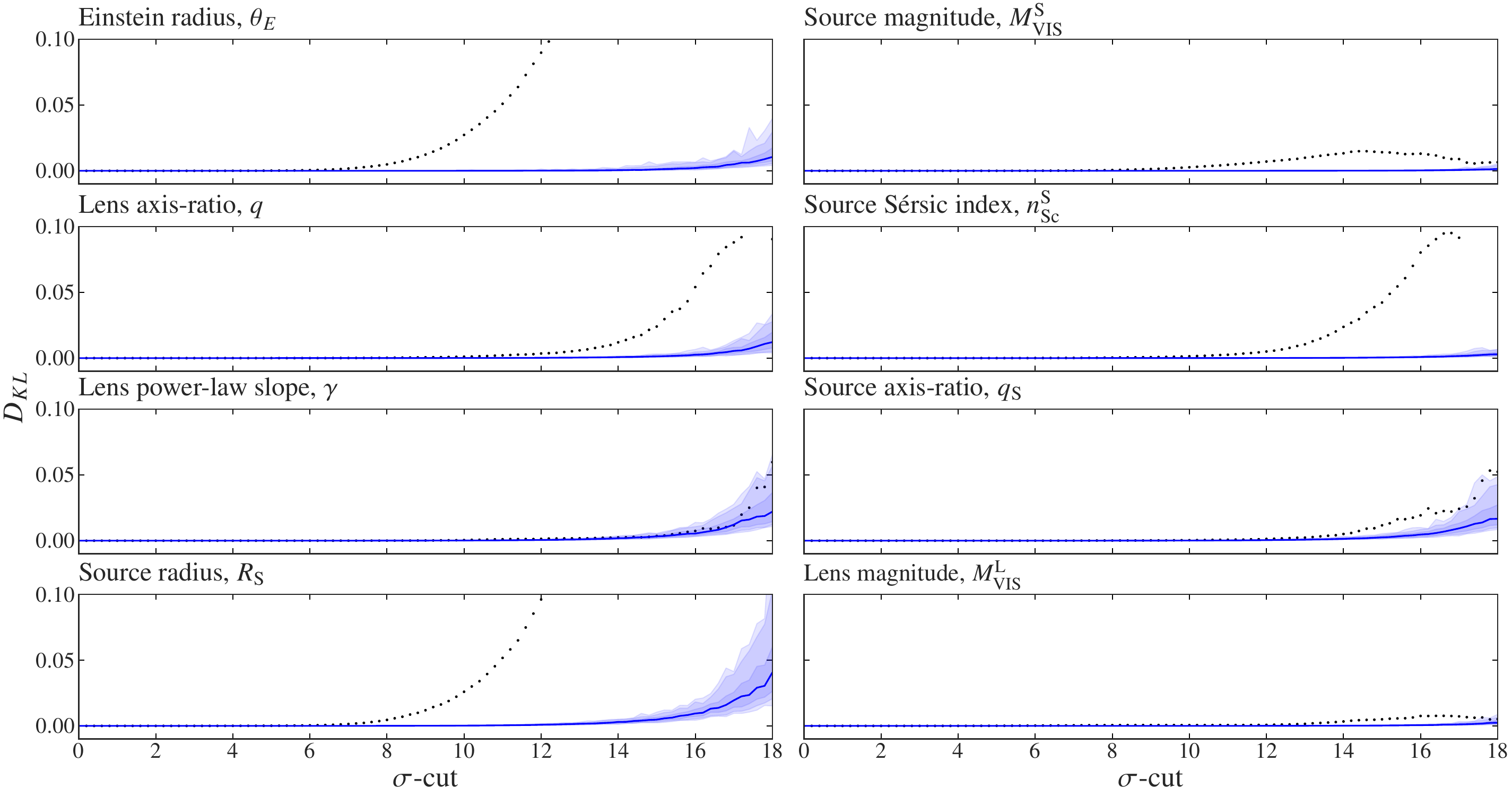}
    \caption{The increase in the Kullback-Leibler Divergence ($D_{KL}$) as the detection significance threshold is increased for the network trained on $\mathcal{D_B}$. The blue lines show the $D_{KL}$ that is due to sampling noise, with the increasingly shaded regions depicting 1, 2 and 3$\sigma$ contours.}
    \label{fig:dkl_B}
\end{figure*}

\begin{figure*}
    \centering
    \includegraphics[scale=0.38]{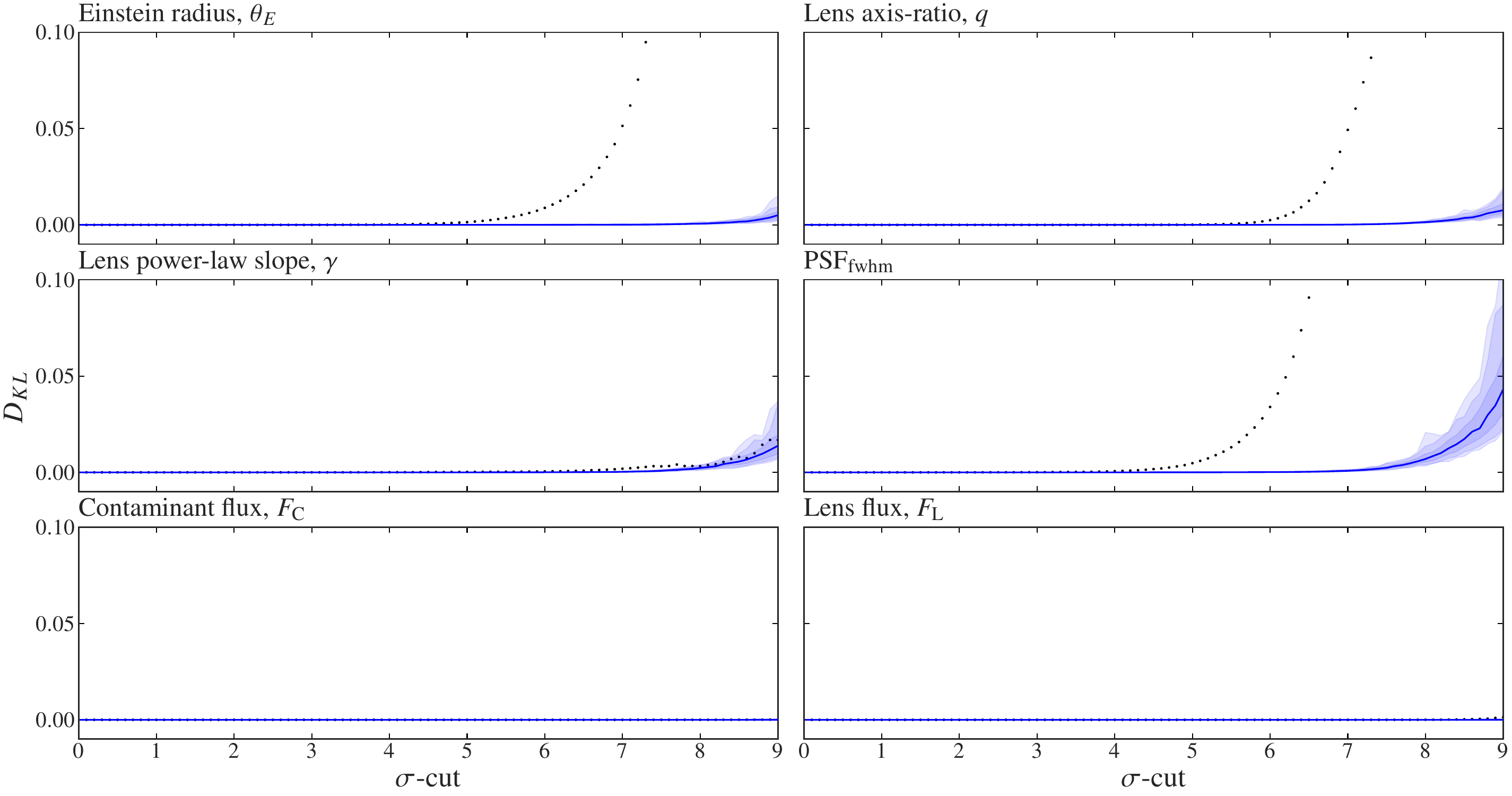}
    \caption{The increase in the Kullback-Leibler Divergence ($D_{KL}$) as the detection significance threshold is increased for the network trained on $\mathcal{D_C}$. The blue lines show the $D_{KL}$ that is due to sampling noise, with the increasingly shaded regions depicting 1, 2 and 3$\sigma$ contours. }
    \label{fig:dkl_C}
\end{figure*}

Our results are shown qualitatively in Figs. \ref{fig:corner-sig-A}, \ref{fig:corner-sig-B} and \ref{fig:corner-sig-C} for the networks trained on $\mathcal{D_A}$, $\mathcal{D_B}$ and $\mathcal{D_C}$, respectively. The contours containing 68 per cent of the mass of the distribution are shown for different detection significance thresholds. As the latter increases, the selected sample begins to deviate from the parent distribution. 

A more quantitative measure is given in terms of the KL-divergence, plotted in Figs. \ref{fig:dkl_A}, \ref{fig:dkl_B} and \ref{fig:dkl_C}, and calculated for a range of detection significance thresholds and for several parameters of interest, as described in Section \ref{sect:KLD}. The blue regions in these figures show the 1, 2 and 3-$\sigma$ contours of the KL-divergence from sampling noise. If a certain parameter is important for a classification, we expect the effect to be captured by the KL-divergence. It should manifest itself as an increase in the KL-divergence between the selected sample and the parent distribution which is much greater than the increase due to sampling noise. We use these variations in the KL-divergence to infer which parameter distributions are biased, and the contour plots to estimate the nature of the bias. Together, they provide a valuable tool to understand the selection function of the three neural networks considered here.

The detection significance threshold at which selection biases become important are high for the neural networks used in this work. However, we urge the reader to keep two important caveats in mind: (i) typically a combination of people and NNs are used to find new lens systems, as the networks find possible targets that are then manually classified \citep[see for example][]{Rojas2023}. The large survey volumes expected necessitate high detection significance cuts in the first pass in order to keep the number of lens candidates that need to be confirmed manually low enough to be manageable. (ii) The actual values of the detection significance when the distributions of the true and selected samples begin to differ strongly is highly dependent on the details of the simulation used to create the training dataset. We have already shown (in Fig. \ref{fig:det-sig-distr}) that the three networks used in this work have distinct distributions of detection significance on their training datasets, even though the differences between them are small. Thus, other lens finder neural networks trained on datasets created with other simulation codes (as will be the case for the networks used on Euclid data, LSST data etc.) will have different detection significance thresholds which are acceptable from a selection bias point of view. The absolute values of the detection significance thresholds at which we see the effect of selection bias manifest for our networks are a relative measure of the importance of each parameter to the network classification, and can only facilitate a comparison between the networks used here. However, we expect the nature and relative strengths of these biases to be a general result for all lens finding CNNs.

From Figs. \ref{fig:corner-sig-A}, \ref{fig:corner-sig-B} and \ref{fig:corner-sig-C}, we see that all three networks show a clear preference for systems with larger Einstein radii. For example, 50 per cent of all ground truth lens systems have Einstein radii $\theta_\mathrm{E}$ $\ge$ 1.25, 0.839 and 0.842 arcsec for $\mathcal{D_A}$, $\mathcal{D_B}$ and $\mathcal{D_C}$ respectively. Beyond 8$\sigma$ however, 50 per cent of the identified lens systems have $\theta_\mathrm{E}$ $\ge$ 0.880 arcsec, and an additional 4$\sigma$ leads to a further increase to $\theta_\mathrm{E}$ $\ge$ 1.04 arcsec for the network trained on $\mathcal{D_A}$. For $\mathcal{D_B}$ and $\mathcal{D_C}$, the same increase in detection significance results in a change in the 50 per cent mark of the selected sample from $\theta_\mathrm{E}$ $\ge$ 0.879 arcsec to $\theta_\mathrm{E}$ $\ge$ 1.04 arcsec and from $\theta_\mathrm{E}$ $\ge$ 0.844 arcsec to $\theta_\mathrm{E}$ $\ge$ 0.975 arcsec respectively. This can be understood intuitively: lensed images that are in general further away from the lens light are easier to de-blend. From the KL-divergence plots, we see that the Einstein radius is the quantity that  most strongly affects the decision of the neural network. Indeed, this is the parameter for which the increase in KL-divergence is the steepest for all three networks. Figs. \ref{fig:dkl_A}, \ref{fig:dkl_B} and \ref{fig:dkl_C} show that the KL-divergence rises sharply at $\approx 8\sigma$ for $\mathcal{D_A}$ and $\mathcal{D_B}$, and at $\approx 6\sigma$ for $\mathcal{D_C}$. Our finding justifies the assumption by \citet{Sonnenfeld2023} to consider $\mathrm{P}_{\mathrm{find}}$ (the probability of finding lens system in a survey) to be solely a function of the Einstein radius.

As the training data is made more complex and realistic, however, other properties of strong gravitational lens systems also become important for the network to correctly identify them as such. The network trained on $\mathcal{D_A}$ shows a small degree of bias at very high detection significance values (> 14$\sigma$) where we see a sensitivity to parameters other than the Einstein radius; namely the lens axis-ratio, source radius, and source and lens magnitudes (as seen in Fig. \ref{fig:dkl_A}). The most important difference between the networks trained on $\mathcal{D_A}$ and $\mathcal{D_B}$ is a bias of the latter in favour of specific source properties. This is related to the fact that $\mathcal{D_A}$ consists of a simpler lens-light model and, therefore, the network does not need to learn complex concepts to distinguish between lensed and un-lensed emission. The more complex lens-light distributions in $\mathcal{D_B}$ cause the network to not only be more sensitive to the lens axis-ratio and the source radius (we see a steep rise in KL-divergence already at > 7$\sigma$ for $R_S$), but also to the S\'ersic index of the source. In particular, it is more likely to select sources with larger radii ($R_\mathrm{S}$), S\'ersic indices ($n_{\mathrm{Sc}}^{\mathrm{S}}$) and axis-ratios ($q_\mathrm{S}$). Specifically, 50 per cent of the selected sample have $R_\mathrm{S}$ $\ge$ 0.194 arcsec, $n_{\mathrm{Sc}}^{\mathrm{S}}$ $\ge$ 2.62 and $q_\mathrm{S}$ $\ge$ 0.758 at a 12-$\sigma$ cut as opposed to the ground truth values of $R_\mathrm{S}$ $\ge$ 0.175 arcsec, $n_{\mathrm{Sc}}^{\mathrm{S}}$ $\ge$ 2.51 and $q_\mathrm{S}$ $\ge$ 0.751. All these properties lead to larger and more concentrated source light distributions, making the lensed arcs more easily distinguishable from the lens light. 

Additionally, all three networks tend to identify as lens systems those with a slightly more elliptical lens mass distribution at very strict detection significance thresholds, (i.e. > 12$\sigma$ for the networks trained on $\mathcal{D_A}$ and $\mathcal{D_B}$). An increase in the detection significance cut from 8 to 12$\sigma$ causes 50 per cent of the selected lens systems to have $q$ $\le$ 0.736 and $q$ $\le$ 0.713 from $q$ $\le$ 0.750 and $q$ $\le$ 0.723, for $\mathcal{D_A}$ and $\mathcal{D_B}$ respectively (with associated true values of $q$ $\le$ 0.750 and $q$ $\le$ 0.723). The KL-divergence plot for $\mathcal{D_C}$ (Fig. \ref{fig:dkl_C}) shows more sensitivity (steep rise in KL-divergence at C > 6$\sigma$) to the lens axis-ratio compared to the extended source lens finders. At a 7-$\sigma$ cut, half of the lens systems have lens axis-ratios $q$ $\le$ 0.670 as opposed to the ground truth values of $q$ $\le$ 0.723. Moreover, this network is more efficient at selecting images with lower $\mathrm{PSF}_{\mathrm{FWHM}}$ values. At lower $\mathrm{PSF}_{\mathrm{FWHM}}$, it becomes easier to distinguish between the lensed point-source and the lens/contaminant galaxies. However, the inclusion of stars as contaminants will make this more difficult, and the preference for lower $\mathrm{PSF}_{\mathrm{FWHM}}$ may be negligible with stars as contaminants. 

For this dataset, we also calculate the KL-divergence as a function of $\sigma$-cut for the flux of the lens $F_{\mathrm{L}}$ and the contaminants, $F_{\mathrm{C}}$, which are proxies for how bright these components are with respect to the source. We find that the KL-divergence is negligible at all values of $\sigma$-cut. This points to the fact that the network has learnt to distinguish between the lens/contaminant light and source light very effectively.

Interestingly, the increase in KL-divergence for the lens power-law slope for all three networks is consistent with the sampling noise, leading us to conclude that there is no preference for mass models of a particular slope.

\section{DISCUSSION}
\label{sect:disc}
Our main findings indicate that the sample of strong gravitational lens systems identified in wide-sky surveys with CNNs are biased with respect to both lens and source properties. This could have important implications for the many scientific applications of strong gravitational lensing. In this section, we discuss the implications of these biases.

The CNN selection function, coupled with the lensing cross-section, will lead to samples of deflectors which are biased towards higher masses. This will effectively limit our prospects of studying the properties of lens galaxies with strong lensing to only the most massive objects. Essentially, the blurring by the PSF produces a lower limit on the Einstein radii of lens systems that can be found in a survey. This effect will be increased by the selection function of the neural networks.

Interestingly, we find no selection bias in favour of specific values of the slope, $\gamma$, of the lens mass density profiles. Assuming that this result is confirmed with other CNNs, we can conclude that the large samples of new lens systems will allow one to measure this parameter and its evolution with redshift. This will provide a valuable probe of  galaxy evolution and feedback models \citep[e.g.][]{Koopmans2006, Mukherjee2021}.

Strong gravitational lensing acts as a cosmic telescope, allowing the study of high-redshift galaxies at higher physical resolution and SNR than usually possible otherwise. We have shown that networks trained on images with complex and more realistic lens light models prefer larger source radii and S\'ersic indices. Sources with these properties are bigger and more concentrated, and thus produce arcs which are easier to distinguish from the lens light. A similar selection bias also occurs from the strong lensing cross-section \citep{Serjeant2012, Hezaveh2012}, and the neural networks will reinforce this effect. Hence, attempts to interpret the properties of lensed galaxies in the context of galaxy evolution models \citep[as for example in][]{Oldham2017a, Oldham2017b, Stacey2021} from the large sample of discovered lenses will need to additionally account for the effect of the neural network selection function.

Time-delay cosmography can be used as an additional probe of the Hubble constant \citep{Refsdal1964, Birrer2020, Birrer2021, Shajib2023}. However, this requires the breaking of the mass-sheet degeneracy \citep{Falco1985} by incorporating independent mass measurements from stellar dynamics. Velocity dispersion measurement uncertainties limit the overall precision on $H_0$ to $\approx$ 10 per cent \citep{Kochanek2020, Schneider2013}. These uncertainties are too large to address the Hubble tension in a meaningful way. To overcome this issue, \citet{Birrer2021} proposed a joint analysis using mass models and kinematic properties of galaxy-galaxy lenses as a prior on the mass profiles for the analysis of the time-delayed galaxy-quasar lenses. This method inherently assumes that two types of strong gravitational lens systems are drawn from the same deflector parent population \citep[see also][]{Sonnenfeld2021}. However, \citet{Sonnenfeld2023} pointed out that in the case of quadruply-imaged quasars, the lens galaxies tend to have larger ellipticities and halo masses for a given stellar mass. Similarly, we find that the neural network trained to find lensed quasars has a preference for higher ellipticity lens profiles as these have wider caustics and thus a higher chance of producing four image systems. More importantly, the networks trained to find lensed extended sources show a much weaker preference for higher ellipticities. Hence, the approach proposed by \citet[][see also \citealt{Gomer2022, Birrer2020}]{Birrer2021} is more likely to introduce additional systematic errors as opposed to constraints. These can be accounted for with an understanding of the selection function of the lens finder used, which can be obtained with an analysis of the type done in this work.
\subsection{Limitations of our work}
\label{sect:lims}
Our image simulations already account for several characteristics of real data that is often ignored in the lens finding literature, like complexity in the lens light models, field galaxies and a variable PSF FWHM. However, the training data used in this work could be further improved by the inclusion of complexities that: (i) make the data more diverse (e.g an elliptical PSF model with varying ellipticities, stars and non-lensed quasars as contaminants, complex source light models, gamma-rays, CCD artefacts) and (ii) make the task of separating lensed emission from source light more difficult (for example ring galaxies). We have seen that complexities like the former require the network to become adept at ignoring contaminants, as with the network trained on $\mathcal{D_C}$. For the latter, we expect a more complex selection function when trained on more realistic data (as with the neural network trained on $\mathcal{D_B}$).
 
We have confined this study to the case of single-band data, thus it is unclear how training with multi-band data will influence the selection function of the neural networks. It is possible that in this scenario, colour related selection biases may be introduced in addition to the ones listed in this paper. Moreover, we have considered only the case of the ResNet18 architecture, as it is often the best performing network for lens finding \citep{Canameras2023}. How the selection function changes for different CNN architectures and for different ML models is an interesting question which is beyond the scope of this work.

Note that our datasets are split evenly between the two classes. A real survey will have far more non-lens than lens images. This is another example of the class imbalance problem in the machine learning literature. A real sample would have at most 1 lens system for every 100 objects, and a neural network trained on a dataset with a class ratio of 1:99 could achieve a 99 per cent accuracy by simply classifying every object as a non-lens. In order to alleviate this issue, the machine learning community has developed many techniques that centre around the theme of artificially making the class ratio closer to 1:1, which can be achieved by over-sampling the minority class or under-sampling the majority class during training. Since we use simulated data, we can circumvent this issue by creating a dataset with an equal number of lens and non-lens systems. Using a realistic class ratio could alter the notions that the networks learn during training. This may further exacerbate the selection effects that we outline in this paper, but we do not study the effect of class imbalance here.
\section{CONCLUSIONS}
\label{sect:conc}
In this paper, we quantified the selection function of the machine learning models that will likely be used to find the strong lens systems in future surveys. We focused our efforts on the ResNet18 architecture trained on three different datasets, and found that the classification task of lens finding leads to a selection bias on the parameters of the identified sample of strong lenses.

This results from the fact that lens finding neural networks are more efficient at finding lenses with certain properties, making them more likely to be above any detection significance threshold chosen for a given survey. Moreover, samples of lenses found by neural networks in the first pass are used as training data for the next iteration of lens finding networks. This might have the effect of further exacerbating the selection bias in each iteration.

We have shown that neural networks are most sensitive to the Einstein radius of the system as they preferentially select strong lens systems with larger values of $\theta_\mathrm{E}$. In addition, the networks are biased towards bigger sources with more concentrated light distributions. Galaxy-quasar lens finding neural networks also show a stronger preference for more elliptical lens mass distributions than those trained to find galaxy-galaxy lens systems. We also find that the networks show no preference for any values of the lens power-law slope.

Lens finding neural networks reinforce the biases introduced by the lensing cross-section. Our results clearly show that an analysis of the selection effects of lens finding neural networks is a key additional step that needs to be incorporated into any systematic attempt to find strong gravitational lenses in upcoming surveys.
\section*{ACKNOWLEDGEMENTS}
AH thanks the Max Planck Computing and Data Facility (MPCDF) for computational resources and support. AH also thanks Daniel Gr\"un, Matteo Guardiani and Philipp Frank for useful insights and discussions. SV thanks the Max Planck Society for support through a Max Planck Lise Meitner Group, and acknowledges funding from the European Research Council (ERC) under the European Union’s Horizon 2020 research and innovation programme (LEDA: grant agreement No 758853).

\section*{DATA AVAILABILITY}

The data used in this paper are available from the corresponding author on request.



\bibliographystyle{mnras}
\bibliography{ms} 

\begin{thebibliography}{}
\makeatletter
\relax
\def\mn@urlcharsother{\let\do\@makeother \do\$\do\&\do\#\do\^\do\_\do\%\do\~}
\def\mn@doi{\begingroup\mn@urlcharsother \@ifnextchar [ {\mn@doi@}
  {\mn@doi@[]}}
\def\mn@doi@[#1]#2{\def\@tempa{#1}\ifx\@tempa\@empty \href
  {http://dx.doi.org/#2} {doi:#2}\else \href {http://dx.doi.org/#2} {#1}\fi
  \endgroup}
\def\mn@eprint#1#2{\mn@eprint@#1:#2::\@nil}
\def\mn@eprint@arXiv#1{\href {http://arxiv.org/abs/#1} {{\tt arXiv:#1}}}
\def\mn@eprint@dblp#1{\href {http://dblp.uni-trier.de/rec/bibtex/#1.xml}
  {dblp:#1}}
\def\mn@eprint@#1:#2:#3:#4\@nil{\def\@tempa {#1}\def\@tempb {#2}\def\@tempc
  {#3}\ifx \@tempc \@empty \let \@tempc \@tempb \let \@tempb \@tempa \fi \ifx
  \@tempb \@empty \def\@tempb {arXiv}\fi \@ifundefined
  {mn@eprint@\@tempb}{\@tempb:\@tempc}{\expandafter \expandafter \csname
  mn@eprint@\@tempb\endcsname \expandafter{\@tempc}}}

\bibitem[\protect\citeauthoryear{{Birrer} \& {Treu}}{{Birrer} \&
  {Treu}}{2021}]{Birrer2021}
{Birrer} S.,  {Treu} T.,  2021, \mn@doi [\aap] {10.1051/0004-6361/202039179},
  \href {https://ui.adsabs.harvard.edu/abs/2021A&A...649A..61B} {649, A61}

\bibitem[\protect\citeauthoryear{{Birrer} et~al.,}{{Birrer}
  et~al.}{2020}]{Birrer2020}
{Birrer} S.,  et~al., 2020, \mn@doi [\aap] {10.1051/0004-6361/202038861}, \href
  {https://ui.adsabs.harvard.edu/abs/2020A&A...643A.165B} {643, A165}

\bibitem[\protect\citeauthoryear{{Ca{\~n}ameras} et~al.,}{{Ca{\~n}ameras}
  et~al.}{2020}]{Canameras2020}
{Ca{\~n}ameras} R.,  et~al., 2020, \mn@doi [\aap]
  {10.1051/0004-6361/202038219}, \href
  {https://ui.adsabs.harvard.edu/abs/2020A&A...644A.163C} {644, A163}

\bibitem[\protect\citeauthoryear{{Ca{\~n}ameras} et~al.,}{{Ca{\~n}ameras}
  et~al.}{2021}]{Canameras2021}
{Ca{\~n}ameras} R.,  et~al., 2021, \mn@doi [\aap]
  {10.1051/0004-6361/202141758}, \href
  {https://ui.adsabs.harvard.edu/abs/2021A&A...653L...6C} {653, L6}

\bibitem[\protect\citeauthoryear{{Ca{\~n}ameras} et~al.,}{{Ca{\~n}ameras}
  et~al.}{2023}]{Canameras2023}
{Ca{\~n}ameras} R.,  et~al., 2023, \mn@doi [arXiv e-prints]
  {10.48550/arXiv.2306.03136}, \href
  {https://ui.adsabs.harvard.edu/abs/2023arXiv230603136C} {p. arXiv:2306.03136}

\bibitem[\protect\citeauthoryear{{Ciotti} \& {Bertin}}{{Ciotti} \&
  {Bertin}}{1999}]{Ciotti1999}
{Ciotti} L.,  {Bertin} G.,  1999, \mn@doi [\aap]
  {10.48550/arXiv.astro-ph/9911078}, \href
  {https://ui.adsabs.harvard.edu/abs/1999A&A...352..447C} {352, 447}

\bibitem[\protect\citeauthoryear{{Collett}}{{Collett}}{2015}]{Collett2015}
{Collett} T.~E.,  2015, \mn@doi [\apj] {10.1088/0004-637X/811/1/20}, \href
  {https://ui.adsabs.harvard.edu/abs/2015ApJ...811...20C} {811, 20}

\bibitem[\protect\citeauthoryear{{Collett} \& {Auger}}{{Collett} \&
  {Auger}}{2014}]{Collett2014}
{Collett} T.~E.,  {Auger} M.~W.,  2014, \mn@doi [\mnras]
  {10.1093/mnras/stu1190}, \href
  {https://ui.adsabs.harvard.edu/abs/2014MNRAS.443..969C} {443, 969}

\bibitem[\protect\citeauthoryear{{Euclid Collaboration} et~al.,}{{Euclid
  Collaboration} et~al.}{2022}]{Euclid2022}
{Euclid Collaboration} et~al., 2022, \mn@doi [\aap]
  {10.1051/0004-6361/202141938}, \href
  {https://ui.adsabs.harvard.edu/abs/2022A&A...662A.112E} {662, A112}

\bibitem[\protect\citeauthoryear{{Falco}, {Gorenstein}  \& {Shapiro}}{{Falco}
  et~al.}{1985}]{Falco1985}
{Falco} E.~E.,  {Gorenstein} M.~V.,   {Shapiro} I.~I.,  1985, \mn@doi [\apjl]
  {10.1086/184422}, \href
  {https://ui.adsabs.harvard.edu/abs/1985ApJ...289L...1F} {289, L1}

\bibitem[\protect\citeauthoryear{{Gilman}, {Birrer}, {Nierenberg}, {Treu}, {Du}
   \& {Benson}}{{Gilman} et~al.}{2020}]{Gilman2020}
{Gilman} D.,  {Birrer} S.,  {Nierenberg} A.,  {Treu} T.,  {Du} X.,   {Benson}
  A.,  2020, \mn@doi [\mnras] {10.1093/mnras/stz3480}, \href
  {https://ui.adsabs.harvard.edu/abs/2020MNRAS.491.6077G} {491, 6077}

\bibitem[\protect\citeauthoryear{{Gomer}, {Sluse}, {Van de Vyvere}, {Birrer}
  \& {Courbin}}{{Gomer} et~al.}{2022}]{Gomer2022}
{Gomer} M.~R.,  {Sluse} D.,  {Van de Vyvere} L.,  {Birrer} S.,   {Courbin} F.,
  2022, \mn@doi [\aap] {10.1051/0004-6361/202244324}, \href
  {https://ui.adsabs.harvard.edu/abs/2022A&A...667A..86G} {667, A86}

\bibitem[\protect\citeauthoryear{{Hezaveh}, {Marrone}  \& {Holder}}{{Hezaveh}
  et~al.}{2012}]{Hezaveh2012}
{Hezaveh} Y.~D.,  {Marrone} D.~P.,   {Holder} G.~P.,  2012, \mn@doi [\apj]
  {10.1088/0004-637X/761/1/20}, \href
  {https://ui.adsabs.harvard.edu/abs/2012ApJ...761...20H} {761, 20}

\bibitem[\protect\citeauthoryear{{Holzschuh}, {O'Riordan}, {Vegetti},
  {Rodriguez-Gomez}  \& {Thuerey}}{{Holzschuh} et~al.}{2022}]{Holzschuh2022}
{Holzschuh} B.~J.,  {O'Riordan} C.~M.,  {Vegetti} S.,  {Rodriguez-Gomez} V.,
  {Thuerey} N.,  2022, \mn@doi [\mnras] {10.1093/mnras/stac1188}, \href
  {https://ui.adsabs.harvard.edu/abs/2022MNRAS.515..652H} {515, 652}

\bibitem[\protect\citeauthoryear{{Hsueh}, {Enzi}, {Vegetti}, {Auger},
  {Fassnacht}, {Despali}, {Koopmans}  \& {McKean}}{{Hsueh}
  et~al.}{2020}]{Hsueh2020}
{Hsueh} J.~W.,  {Enzi} W.,  {Vegetti} S.,  {Auger} M.~W.,  {Fassnacht} C.~D.,
  {Despali} G.,  {Koopmans} L.~V.~E.,   {McKean} J.~P.,  2020, \mn@doi [\mnras]
  {10.1093/mnras/stz3177}, \href
  {https://ui.adsabs.harvard.edu/abs/2020MNRAS.492.3047H} {492, 3047}

\bibitem[\protect\citeauthoryear{{Huang} et~al.,}{{Huang}
  et~al.}{2020}]{Huang2020}
{Huang} X.,  et~al., 2020, \mn@doi [\apj] {10.3847/1538-4357/ab7ffb}, \href
  {https://ui.adsabs.harvard.edu/abs/2020ApJ...894...78H} {894, 78}

\bibitem[\protect\citeauthoryear{{Huang} et~al.,}{{Huang}
  et~al.}{2021}]{Huang2021}
{Huang} X.,  et~al., 2021, \mn@doi [\apj] {10.3847/1538-4357/abd62b}, \href
  {https://ui.adsabs.harvard.edu/abs/2021ApJ...909...27H} {909, 27}

\bibitem[\protect\citeauthoryear{{Kochanek}}{{Kochanek}}{2020}]{Kochanek2020}
{Kochanek} C.~S.,  2020, \mn@doi [\mnras] {10.1093/mnras/staa344}, \href
  {https://ui.adsabs.harvard.edu/abs/2020MNRAS.493.1725K} {493, 1725}

\bibitem[\protect\citeauthoryear{{Koopmans}, {Treu}, {Bolton}, {Burles}  \&
  {Moustakas}}{{Koopmans} et~al.}{2006}]{Koopmans2006}
{Koopmans} L. V.~E.,  {Treu} T.,  {Bolton} A.~S.,  {Burles} S.,   {Moustakas}
  L.~A.,  2006, \mn@doi [\apj] {10.1086/505696}, \href
  {https://ui.adsabs.harvard.edu/abs/2006ApJ...649..599K} {649, 599}

\bibitem[\protect\citeauthoryear{{Lanusse}, {Ma}, {Li}, {Collett}, {Li},
  {Ravanbakhsh}, {Mandelbaum}  \& {P{\'o}czos}}{{Lanusse}
  et~al.}{2018}]{Lanusse2018}
{Lanusse} F.,  {Ma} Q.,  {Li} N.,  {Collett} T.~E.,  {Li} C.-L.,  {Ravanbakhsh}
  S.,  {Mandelbaum} R.,   {P{\'o}czos} B.,  2018, \mn@doi [\mnras]
  {10.1093/mnras/stx1665}, \href
  {https://ui.adsabs.harvard.edu/abs/2018MNRAS.473.3895L} {473, 3895}

\bibitem[\protect\citeauthoryear{{McKean} et~al.,}{{McKean}
  et~al.}{2015}]{McKean2015}
{McKean} J.,  et~al., 2015, in Advancing Astrophysics with the Square Kilometre
  Array (AASKA14). p.~84 (\mn@eprint {arXiv} {1502.03362}),
  \mn@doi{10.22323/1.215.0084}

\bibitem[\protect\citeauthoryear{{Mukherjee}, {Koopmans}, {Metcalf}, {Tortora},
  {Schaller}, {Schaye}, {Vernardos}  \& {Bellagamba}}{{Mukherjee}
  et~al.}{2021}]{Mukherjee2021}
{Mukherjee} S.,  {Koopmans} L. V.~E.,  {Metcalf} R.~B.,  {Tortora} C.,
  {Schaller} M.,  {Schaye} J.,  {Vernardos} G.,   {Bellagamba} F.,  2021,
  \mn@doi [\mnras] {10.1093/mnras/stab693}, \href
  {https://ui.adsabs.harvard.edu/abs/2021MNRAS.504.3455M} {504, 3455}

\bibitem[\protect\citeauthoryear{{O'Riordan}, {Warren}  \&
  {Mortlock}}{{O'Riordan} et~al.}{2019}]{O'Riordan2019}
{O'Riordan} C.~M.,  {Warren} S.~J.,   {Mortlock} D.~J.,  2019, \mn@doi [\mnras]
  {10.1093/mnras/stz1603}, \href
  {https://ui.adsabs.harvard.edu/abs/2019MNRAS.487.5143O} {487, 5143}

\bibitem[\protect\citeauthoryear{{O'Riordan}, {Despali}, {Vegetti}, {Lovell}
  \& {Molin{\'e}}}{{O'Riordan} et~al.}{2023}]{O'Riordan2023}
{O'Riordan} C.~M.,  {Despali} G.,  {Vegetti} S.,  {Lovell} M.~R.,
  {Molin{\'e}} {\'A}.,  2023, \mn@doi [\mnras] {10.1093/mnras/stad650}, \href
  {https://ui.adsabs.harvard.edu/abs/2023MNRAS.521.2342O} {521, 2342}

\bibitem[\protect\citeauthoryear{{Oldham} et~al.,}{{Oldham}
  et~al.}{2017a}]{Oldham2017a}
{Oldham} L.,  et~al., 2017a, \mn@doi [\mnras] {10.1093/mnras/stw2832}, \href
  {https://ui.adsabs.harvard.edu/abs/2017MNRAS.465.3185O} {465, 3185}

\bibitem[\protect\citeauthoryear{{Oldham}, {Auger}, {Fassnacht}, {Treu},
  {Koopmans}, {Lagattuta}, {McKean}  \& {Vegetti}}{{Oldham}
  et~al.}{2017b}]{Oldham2017b}
{Oldham} L.,  {Auger} M.,  {Fassnacht} C.~D.,  {Treu} T.,  {Koopmans} L.~V.~E.,
   {Lagattuta} D.,  {McKean} J.,   {Vegetti} S.,  2017b, \mn@doi [\mnras]
  {10.1093/mnras/stx1127}, \href
  {https://ui.adsabs.harvard.edu/abs/2017MNRAS.470.3497O} {470, 3497}

\bibitem[\protect\citeauthoryear{{Petrillo} et~al.,}{{Petrillo}
  et~al.}{2017}]{Petrillo2017}
{Petrillo} C.~E.,  et~al., 2017, \mn@doi [\mnras] {10.1093/mnras/stx2052},
  \href {https://ui.adsabs.harvard.edu/abs/2017MNRAS.472.1129P} {472, 1129}

\bibitem[\protect\citeauthoryear{{Petrillo} et~al.,}{{Petrillo}
  et~al.}{2019a}]{Petrillo2019a}
{Petrillo} C.~E.,  et~al., 2019a, \mn@doi [\mnras] {10.1093/mnras/sty2683},
  \href {https://ui.adsabs.harvard.edu/abs/2019MNRAS.482..807P} {482, 807}

\bibitem[\protect\citeauthoryear{{Petrillo} et~al.,}{{Petrillo}
  et~al.}{2019b}]{Petrillo2019b}
{Petrillo} C.~E.,  et~al., 2019b, \mn@doi [\mnras] {10.1093/mnras/stz189},
  \href {https://ui.adsabs.harvard.edu/abs/2019MNRAS.484.3879P} {484, 3879}

\bibitem[\protect\citeauthoryear{{Pourrahmani}, {Nayyeri}  \&
  {Cooray}}{{Pourrahmani} et~al.}{2018}]{Pourrahmani2018}
{Pourrahmani} M.,  {Nayyeri} H.,   {Cooray} A.,  2018, \mn@doi [\apj]
  {10.3847/1538-4357/aaae6a}, \href
  {https://ui.adsabs.harvard.edu/abs/2018ApJ...856...68P} {856, 68}

\bibitem[\protect\citeauthoryear{{Refsdal}}{{Refsdal}}{1964}]{Refsdal1964}
{Refsdal} S.,  1964, \mn@doi [\mnras] {10.1093/mnras/128.4.307}, \href
  {https://ui.adsabs.harvard.edu/abs/1964MNRAS.128..307R} {128, 307}

\bibitem[\protect\citeauthoryear{{Rezaei}, {McKean}, {Biehl}, {de Roo}  \&
  {Lafontaine}}{{Rezaei} et~al.}{2022}]{Rezaei2022}
{Rezaei} S.,  {McKean} J.~P.,  {Biehl} M.,  {de Roo} W.,   {Lafontaine} A.,
  2022, \mn@doi [\mnras] {10.1093/mnras/stac2078}, \href
  {https://ui.adsabs.harvard.edu/abs/2022MNRAS.517.1156R} {517, 1156}

\bibitem[\protect\citeauthoryear{{Ritondale}, {Vegetti}, {Despali}, {Auger},
  {Koopmans}  \& {McKean}}{{Ritondale} et~al.}{2019}]{Ritondale2019}
{Ritondale} E.,  {Vegetti} S.,  {Despali} G.,  {Auger} M.~W.,  {Koopmans}
  L.~V.~E.,   {McKean} J.~P.,  2019, \mn@doi [\mnras] {10.1093/mnras/stz464},
  \href {https://ui.adsabs.harvard.edu/abs/2019MNRAS.485.2179R} {485, 2179}

\bibitem[\protect\citeauthoryear{{Rizzo}, {Vegetti}, {Fraternali}, {Stacey}  \&
  {Powell}}{{Rizzo} et~al.}{2021}]{Rizzo2021}
{Rizzo} F.,  {Vegetti} S.,  {Fraternali} F.,  {Stacey} H.~R.,   {Powell} D.,
  2021, \mn@doi [\mnras] {10.1093/mnras/stab2295}, \href
  {https://ui.adsabs.harvard.edu/abs/2021MNRAS.507.3952R} {507, 3952}

\bibitem[\protect\citeauthoryear{{Rodriguez-Gomez} et~al.,}{{Rodriguez-Gomez}
  et~al.}{2019}]{Rodriguez-Gomez2019}
{Rodriguez-Gomez} V.,  et~al., 2019, \mn@doi [\mnras] {10.1093/mnras/sty3345},
  \href {https://ui.adsabs.harvard.edu/abs/2019MNRAS.483.4140R} {483, 4140}

\bibitem[\protect\citeauthoryear{{Rojas} et~al.,}{{Rojas}
  et~al.}{2023}]{Rojas2023}
{Rojas} K.,  et~al., 2023, \mn@doi [\mnras] {10.1093/mnras/stad1680}, \href
  {https://ui.adsabs.harvard.edu/abs/2023MNRAS.523.4413R} {523, 4413}

\bibitem[\protect\citeauthoryear{{Schneider} \& {Sluse}}{{Schneider} \&
  {Sluse}}{2013}]{Schneider2013}
{Schneider} P.,  {Sluse} D.,  2013, \mn@doi [\aap]
  {10.1051/0004-6361/201321882}, \href
  {https://ui.adsabs.harvard.edu/abs/2013A&A...559A..37S} {559, A37}

\bibitem[\protect\citeauthoryear{{Scott}}{{Scott}}{2015}]{Scott2015}
{Scott} D.~W.,  2015, {Multivariate Density Estimation: Theory, Practice, and
  Visualization}

\bibitem[\protect\citeauthoryear{{Serjeant}}{{Serjeant}}{2012}]{Serjeant2012}
{Serjeant} S.,  2012, \mn@doi [\mnras] {10.1111/j.1365-2966.2012.20761.x},
  \href {https://ui.adsabs.harvard.edu/abs/2012MNRAS.424.2429S} {424, 2429}

\bibitem[\protect\citeauthoryear{{S{\'e}rsic}}{{S{\'e}rsic}}{1963}]{Sersic1963}
{S{\'e}rsic} J.~L.,  1963, Boletin de la Asociacion Argentina de Astronomia La
  Plata Argentina, \href
  {https://ui.adsabs.harvard.edu/abs/1963BAAA....6...41S} {6, 41}

\bibitem[\protect\citeauthoryear{{Shajib} et~al.,}{{Shajib}
  et~al.}{2023}]{Shajib2023}
{Shajib} A.~J.,  et~al., 2023, \mn@doi [\aap] {10.1051/0004-6361/202345878},
  \href {https://ui.adsabs.harvard.edu/abs/2023A&A...673A...9S} {673, A9}

\bibitem[\protect\citeauthoryear{{Sonnenfeld}}{{Sonnenfeld}}{2021}]{Sonnenfeld2021}
{Sonnenfeld} A.,  2021, \mn@doi [\aap] {10.1051/0004-6361/202142062}, \href
  {https://ui.adsabs.harvard.edu/abs/2021A&A...656A.153S} {656, A153}

\bibitem[\protect\citeauthoryear{{Sonnenfeld}}{{Sonnenfeld}}{2022}]{Sonnenfeld2022}
{Sonnenfeld} A.,  2022, \mn@doi [\aap] {10.1051/0004-6361/202142301}, \href
  {https://ui.adsabs.harvard.edu/abs/2022A&A...659A.132S} {659, A132}

\bibitem[\protect\citeauthoryear{{Sonnenfeld}, {Jaelani}, {Chan}, {More},
  {Suyu}, {Wong}, {Oguri}  \& {Lee}}{{Sonnenfeld}
  et~al.}{2019}]{Sonnenfeld2019}
{Sonnenfeld} A.,  {Jaelani} A.~T.,  {Chan} J.,  {More} A.,  {Suyu} S.~H.,
  {Wong} K.~C.,  {Oguri} M.,   {Lee} C.-H.,  2019, \mn@doi [\aap]
  {10.1051/0004-6361/201935743}, \href
  {https://ui.adsabs.harvard.edu/abs/2019A&A...630A..71S} {630, A71}

\bibitem[\protect\citeauthoryear{{Sonnenfeld}, {Li}, {Despali}, {Shajib}  \&
  {Taylor}}{{Sonnenfeld} et~al.}{2023}]{Sonnenfeld2023}
{Sonnenfeld} A.,  {Li} S.-S.,  {Despali} G.,  {Shajib} A.~J.,   {Taylor} E.~N.,
   2023, \mn@doi [arXiv e-prints] {10.48550/arXiv.2301.13230}, \href
  {https://ui.adsabs.harvard.edu/abs/2023arXiv230113230S} {p. arXiv:2301.13230}

\bibitem[\protect\citeauthoryear{{Springel} et~al.,}{{Springel}
  et~al.}{2018}]{Springel2018}
{Springel} V.,  et~al., 2018, \mn@doi [\mnras] {10.1093/mnras/stx3304}, \href
  {https://ui.adsabs.harvard.edu/abs/2018MNRAS.475..676S} {475, 676}

\bibitem[\protect\citeauthoryear{{Stacey} et~al.,}{{Stacey}
  et~al.}{2021}]{Stacey2021}
{Stacey} H.~R.,  et~al., 2021, \mn@doi [\mnras] {10.1093/mnras/staa3433}, \href
  {https://ui.adsabs.harvard.edu/abs/2021MNRAS.500.3667S} {500, 3667}

\bibitem[\protect\citeauthoryear{{Vegetti}, {Despali}, {Lovell}  \&
  {Enzi}}{{Vegetti} et~al.}{2018}]{Vegetti2018}
{Vegetti} S.,  {Despali} G.,  {Lovell} M.~R.,   {Enzi} W.,  2018, \mn@doi
  [\mnras] {10.1093/mnras/sty2393}, \href
  {https://ui.adsabs.harvard.edu/abs/2018MNRAS.481.3661V} {481, 3661}

\makeatother
\end{thebibliography}




\appendix

\section{Appendix}
\subsection{From logits to detection significance}
\label{sect:logits}

The networks were trained for a binary classification task, i.e they produce one 2-D vector corresponding to the logits assigned for each class per image. This is encapsulated in the following equations:
\begin{equation}
    f(X_i) = Y_i = (Y_{\mathrm{NL}, i}, Y_{\mathrm{L}, i})\,,
\end{equation}
Where $x_i$ is the input to the network ($i$-th instance of the test dataset) and $f()$ is the mapping learnt by the network. The outputs correspond to a logit value for each of the two possible classes, \emph{lens} and \emph{non-lens}. $Y_{\mathrm{NL}, i}$ and $Y_{\mathrm{L}, i}$ are the non-lens and lens logits respectively, and are typically passed through a softmax function which transforms the network's outputs from logit space to prediction space. The softmax function is defined by:
\begin{equation}
    \mathrm{softmax} (z_i) = \frac{e^{z_i} }{\sum_{j=0}^N e^{z_j}}\,,
\end{equation}
Where $z$ is the vector corresponding to the output of the final fully-connected layer, and $K$ is the number of classes the network is trained to classify. In our case, the final prediction probability for a given input being a lens is:
\begin{equation}
    p_i(\mathrm{L}) = \frac{e^{Y_{\mathrm{L}, i}}}{e^{Y_{\mathrm{NL}, i}}+e^{Y_{\mathrm{L}, i}}}\,,
\end{equation}
and 
\begin{equation}
    p_i(\mathrm{NL}) = \frac{e^{Y_{\mathrm{NL}, i}}}{e^{Y_{\mathrm{NL}, i}}+e^{Y_{\mathrm{L}, i}}} = 1 - p_i(\mathrm{L})\,,
\end{equation}
where $\mathrm{L}$ and $\mathrm{NL}$ are the lens and non-lens classes respectively.

The inverse error function provides a measure of how close an input is to $1.0$. The function has a null-value at $0$ and approaches infinity asymptotically at $1.0$, which can be interpreted as a detection significance. Thus, a \emph{lens} image that produces a probability of being a lens of exactly $1.0$ is given an infinite detection significance, and one that produces a $0.99$ is given a $3 \sigma$ detection confidence. Between the probabilities of $0.99$ and $1.0$, the network's decision confidence is easier to interpret if one makes a conversion through the inverse error function.

However, conversion from logits to detection significance using the inverse error function is non-trivial as one runs into numerical precision issues due to the asymptotic behaviour of the function as it approaches $1.0$. Essentially, high-precision floating point calculations need to be performed in order to make this conversion. Interestingly, we find that converting from logit space directly to detection significance (without the intermediate conversion to probability space via the softmax function) can be done using the relation:
\begin{equation}
    C = \sqrt{2 Y_{\mathrm{L}, i}}\,,
\end{equation}
Where $C$ is the network's lens detection significance.
\begin{figure*}
    \centering
    \includegraphics[scale=0.7]{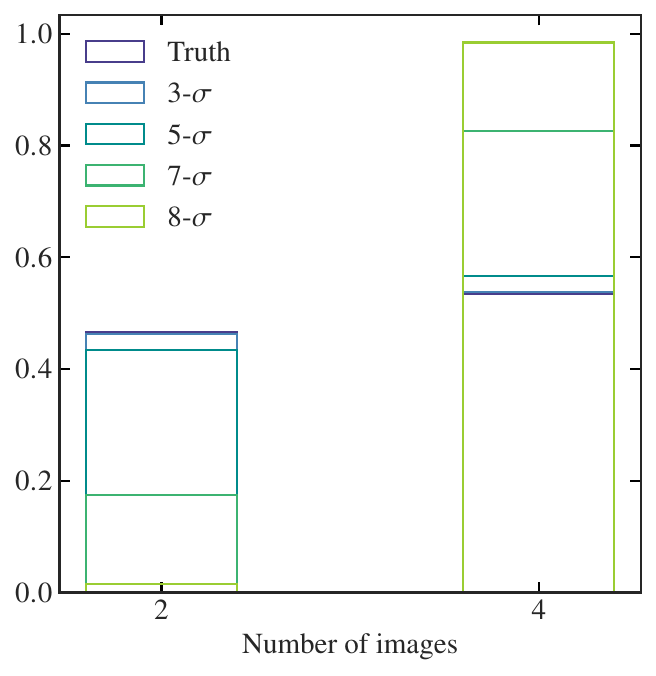}
    \includegraphics[scale=0.7]{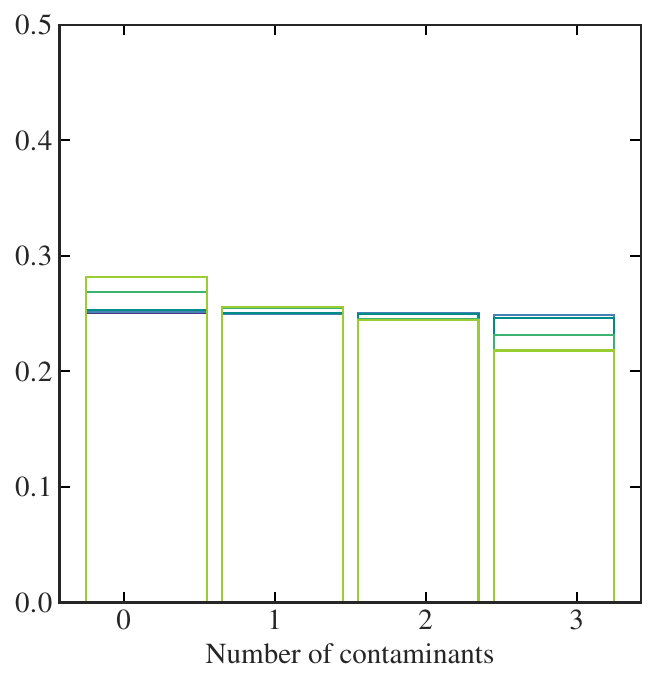}
    \caption{Left: Distribution of the number of images of the source at different detection significance thresholds for the network trained on $\mathcal{D_C}$. Right: Distribution of the number of contaminants (field galaxies) in the image at different detection significance thresholds for the network trained on $\mathcal{D_C}$.}
    \label{fig:num_imgs_conts}
\end{figure*}
\subsection{Number of images and contaminants}
It is ensured that the number of quadruply- and doubly-lensed quasars in $\mathcal{D_C}$ are roughly equal (in reality, we would expect more doubles than quads). There is also a negligible fraction of three-image systems. As the detection significance is increased, we see in the left panel of Fig. \ref{fig:num_imgs_conts} that the network very strongly prefers quadruply-imaged quasars as opposed to doubly-imaged ones.

In the testing dataset, the number of contaminants in each image is sampled uniformly between 0 and 4. The right panel of Fig. \ref{fig:num_imgs_conts} shows the distributions of the number of contaminants in the image at different detection significance thresholds. The plot indicates a subtle preference for images with fewer field galaxies as these may make the separation of the lensed point images from the rest of the field more complex. 

\bsp	
\label{lastpage}
\end{document}